\documentclass{aa}
\input epsf
\begin{document}
\thesaurus{23  
	   (11.09.1 
	    11.16.1 
	    11.19.2 
	    11.19.4 
	    03.20.1) 
	    }
\title{Young massive star clusters in nearby galaxies
  \thanks{Based on observations made with the Nordic Optical Telescope,
	  operated on the island of La Palma jointly by Denmark, Finland,
          Iceland, Norway, and Sweden, in the Spanish Observatorio del 
          Roque de los Muchachos of the Instituto de Astrofisica de Canarias,
	  and with the Danish 1.5-m telescope at ESO, La Silla, Chile.}
}
\subtitle{II. Software tools, data reductions and cluster sizes} 
\author{S. S. Larsen}
\institute{Copenhagen University Astronomical Observatory,
Juliane Maries Vej 32, 2100 Copenhagen {\O}, Denmark \\
email: soeren@astro.ku.dk
}
\date{Received ...; accepted ...}
\maketitle
\markboth{Young massive star clusters in nearby galaxies}{}
\begin{abstract}
  \footnote{Table \ref{tab:clusters} is also available in electronic form at
	    the CDS via http://cdsweb.u-strasbg.fr/Abstract.html}

  A detailed description is given of the data analysis leading to the
discovery of young massive star clusters (YMCs) in a sample of 21 nearby 
galaxies.  A new useful tool, {\bf ishape}, for the derivation of intrinsic 
shape parameters of compact objects is presented, and some test
results are shown. Completeness tests for the cluster samples are
discussed, and {\bf ishape} is used to estimate cluster sizes. 
Half-light radii of 0 - 20 pc are derived for clusters in 2 of the most 
cluster-rich galaxies, NGC~1313 and NGC~5236, which is within the range 
spanned by globular clusters in the Milky Way and by YMCs in the LMC and 
some starburst galaxies.  Photometric data for all clusters, along with 
positions and estimated half-light radii, are tabulated.  
\end{abstract}

\section{Introduction}

  This paper is the second in a series describing the observations of 
``young massive star clusters'' (YMCs) in a number of nearby spiral and
irregular galaxies. A general outline of the project was given by 
Larsen \& Richtler (\cite{lr1999}, Paper1), and some first results were 
presented 
there. In this paper the data reductions are discussed in more detail, with
special emphasis on two software tools developed as part of the project, 
which may be of general use to the astronomical community.  These tools 
are {\bf mksynth}, a programme which generates synthetic images with 
realistic noise characteristics, and {\bf ishape} which derives intrinsic 
shape parameters for compact objects in a digital image by convolving
an analytic model profile with the PSF and matching the result to the
observed profile.

  By YMCs we mean objects of the same type as e.g. the cluster NGC~1866 in 
the Large Magellanic Cloud (Fischer et al. \cite{fischer1992}). They are 
characterised by being similar (with respect to mass and morphology) to 
globular clusters seen in the Milky Way, but of much lower ages - in fact, 
the lower age limit of the clusters in our sample is set only by the fact 
that we avoid objects located within HII regions. 

  The paper is organised as follows: The observations are described in 
Sect.~\ref{sec:obs}, and Sect.~\ref{sec:red} is a discussion of the basic 
data reduction procedure including the standard calibrations.  In 
Sect.~\ref{sec:tools} the tools {\bf mksynth} and {\bf ishape} are 
introduced, and Sect.~\ref{sec:compl} gives a discussion of the completeness 
corrections used in Paper1. Finally, in Sect.~\ref{sec:resolv} {\bf ishape} 
is used to derive radii of the clusters, and some implications of the 
results are discussed.

\section{Observations}
\label{sec:obs} 

\begin{figure*}
\begin{minipage}{16cm}
\begin{minipage}{5cm}
NGC 45\\
 \epsfxsize=5cm
 \epsfbox{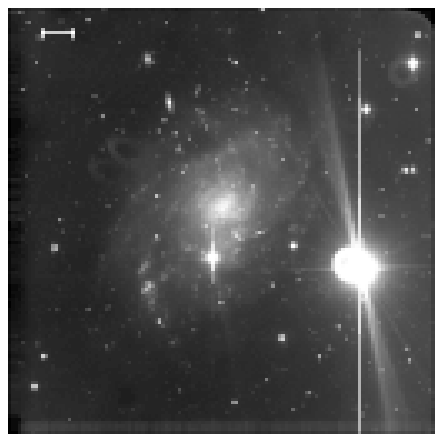} \\
NGC 247\\
 \epsfxsize=5cm
 \epsfbox{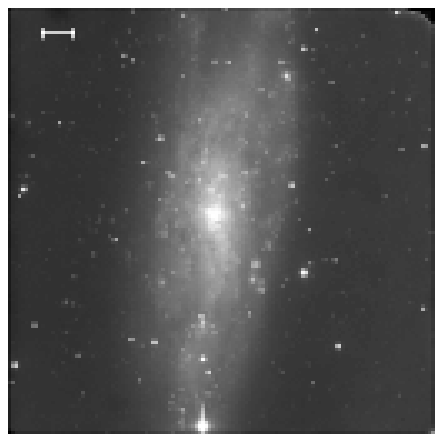} \\
NGC 300\\
 \epsfxsize=5cm
 \epsfbox{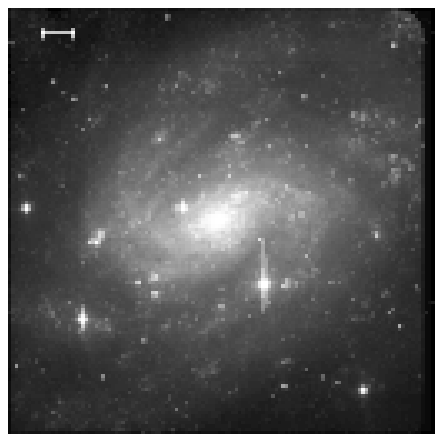} \\
NGC 628\\
 \epsfxsize=5cm
 \epsfbox{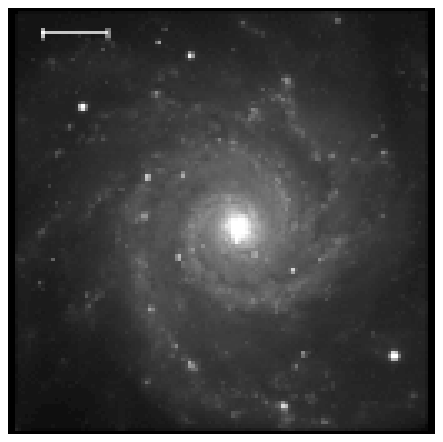} \\
\end{minipage}
\begin{minipage}{5cm}
NGC 1156\\
 \epsfxsize=5cm
 \epsfbox{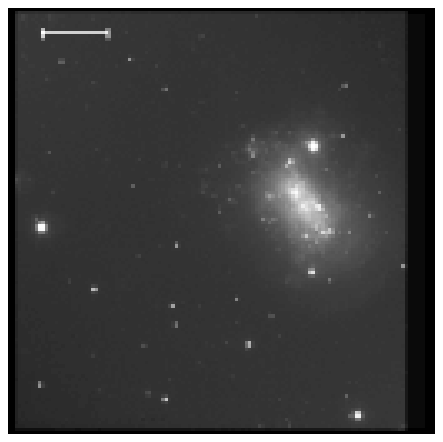}\\
NGC 1313\\
 \epsfxsize=5cm
 \epsfbox{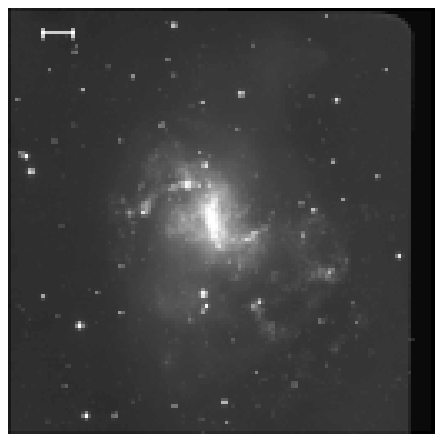}\\
NGC 1493\\
 \epsfxsize=5cm
 \epsfbox{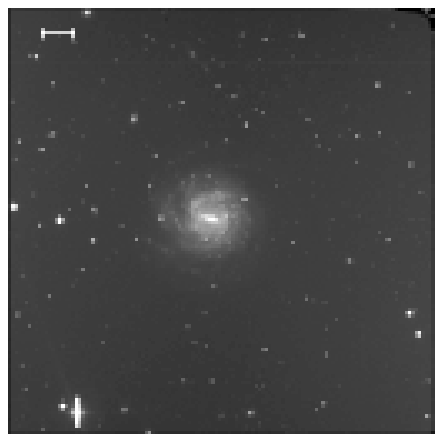} \\
NGC 2403\\
 \epsfxsize=5cm
 \epsfbox{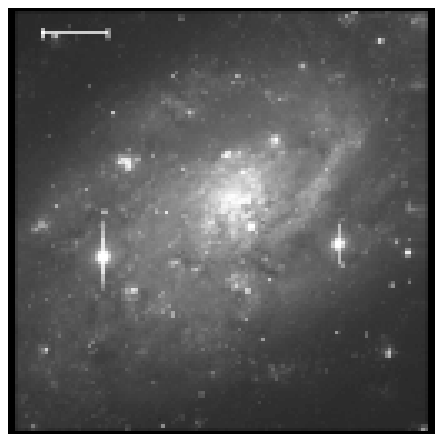} \\
\end{minipage}
\begin{minipage}{5cm}
NGC 2835 \\
 \epsfxsize=5cm
 \epsfbox{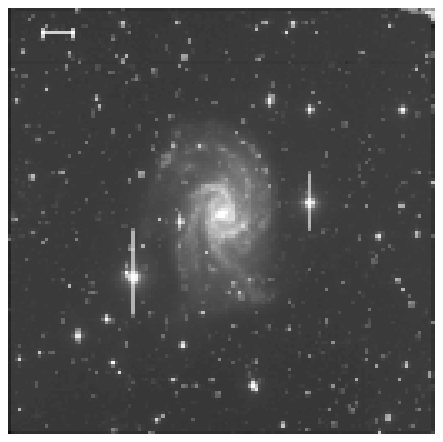} \\
NGC 2997\\
 \epsfxsize=5cm
 \epsfbox{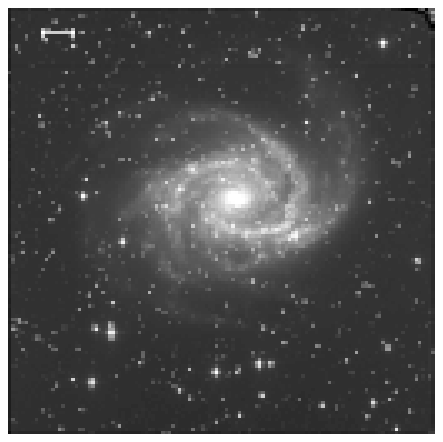} \\
NGC 3184\\
 \epsfxsize=5cm
 \epsfbox{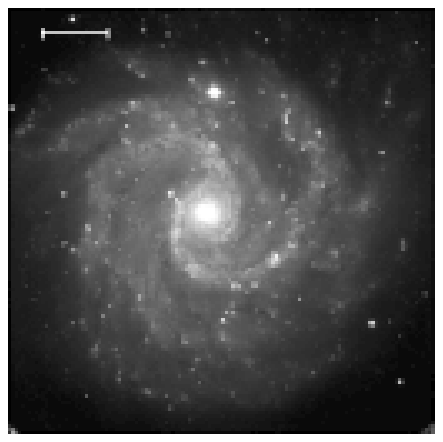} \\
NGC 3621\\
 \epsfxsize=5cm
 \epsfbox{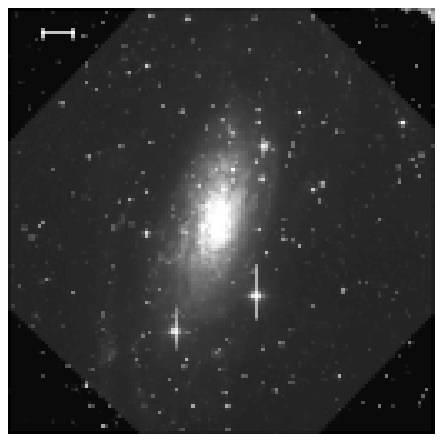} \\
\end{minipage}
\end{minipage}
\caption{
  \mbox{$V$-band} images of the galaxies. The bar in the upper left corner
 of each image corresponds to 1 arcminute. North is up and east to the left
 in all images.
 \label{fig:im1}
}
\end{figure*}

\begin{figure*}
\begin{minipage}{16cm}
\begin{minipage}{5cm}
NGC 4395\\
 \epsfxsize=5cm
 \epsfbox{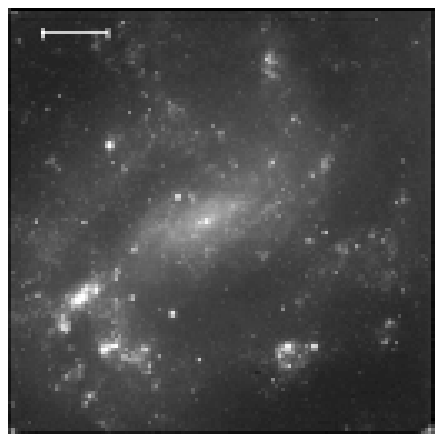} \\
NGC 5204\\
 \epsfxsize=5cm
 \epsfbox{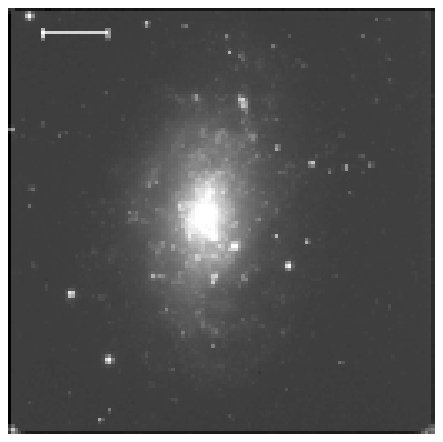} \\
NGC 5236\\
 \epsfxsize=5cm
 \epsfbox{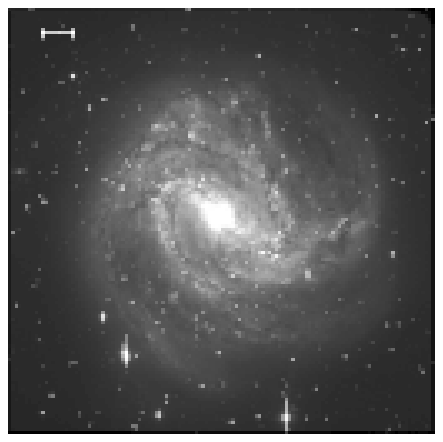} \\
\end{minipage}
\begin{minipage}{5cm}
NGC 5585\\
 \epsfxsize=5cm
 \epsfbox{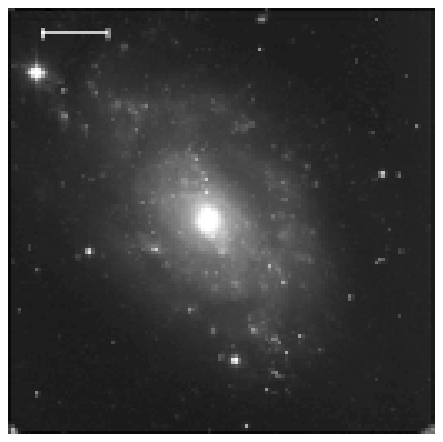} \\
NGC 6744\\
 \epsfxsize=5cm
 \epsfbox{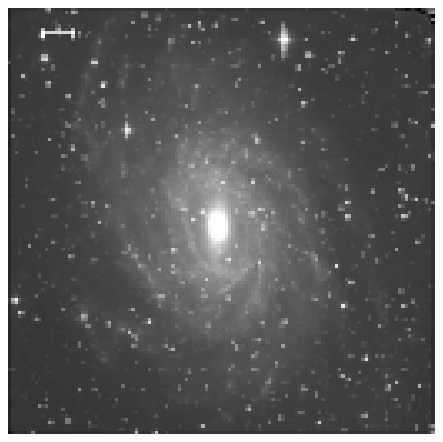} \\
NGC 6946\\
 \epsfxsize=5cm
 \epsfbox{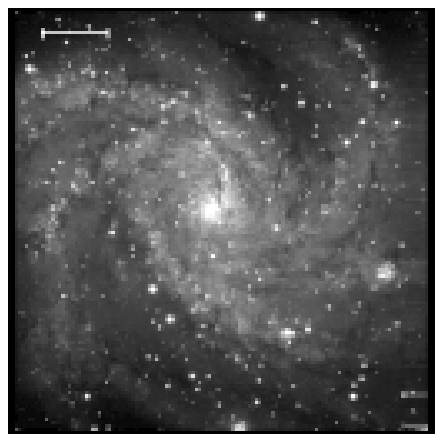} \\
\end{minipage}
\begin{minipage}{5cm}
NGC 7424\\
 \epsfxsize=5cm
 \epsfbox{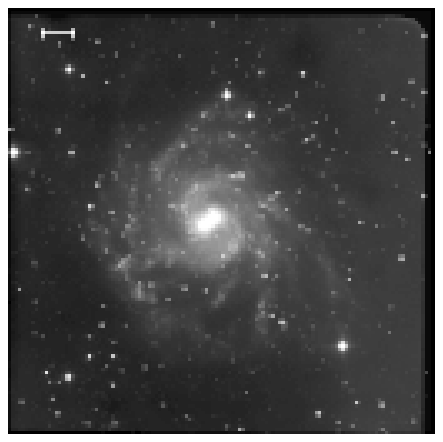} \\
NGC 7741\\
 \epsfxsize=5cm
 \epsfbox{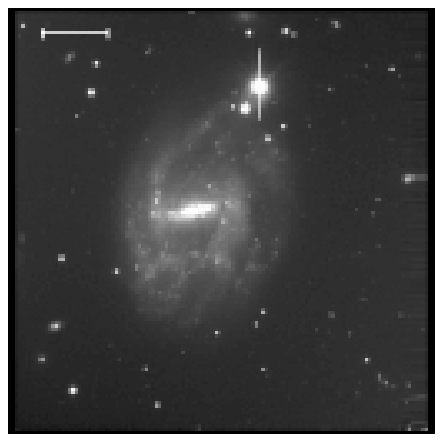} \\
NGC 7793\\
 \epsfxsize=5cm
 \epsfbox{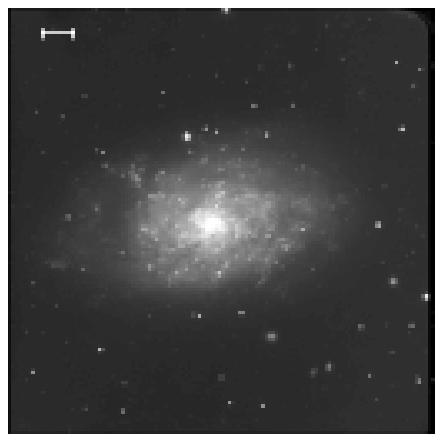} \\
\end{minipage}
\end{minipage}
\caption{
  See the caption to Fig.~\ref{fig:im1}.
  \label{fig:im2}
}
\end{figure*}

  Observations of 21 nearby galaxies, listed in Table 1 of Paper1,  were 
carried out with the Danish 1.54 m. telescope and DFOSC (Danish Faint
Object Spectrograph and Camera) at the European Southern Observatory (ESO) 
at La Silla, Chile, and with the 2.56 m. Nordic Optical Telescope and ALFOSC 
(a DFOSC twin instrument), situated at La Palma, Canary Islands. The data 
consist of CCD images in the filters U,B,V,R,I and H$\alpha$. In the filters 
BVRI and H$\alpha$ we typically made 3 exposures per galaxy of 5 minutes 
each, and 3 exposures of 20 minutes each in the U band. V-band images of
the galaxies are shown in Figs.~\ref{fig:im1} and \ref{fig:im2}.

  Both the ALFOSC
and DFOSC were equipped with thinned, backside-illuminated 2 K$^2$ 
Loral-Lesser CCDs. The pixel scale in the ALFOSC is 0.189\arcsec /pixel and
the scale in the DFOSC is 0.40\arcsec /pixel, giving field sizes
of $6.5\arcmin \times 6.5\arcmin$ and $13.7\arcmin \times 13.7\arcmin$, 
respectively.

  The seeing during the observations at La Silla was typically around
1.5 arcseconds as measured on V-band frames. On the images taken with the
NOT the seeing was usually about 0.8 arcseconds.  All observations used in 
this paper were conducted under photometric conditions.

  During each observing run, photometric standard stars in the 
Landolt~(\cite{landolt1992}) 
fields were observed for calibration of the photometry. Care was taken to 
include standard stars over as a wide a range in colours as possible, usually
from $B-V \approx -0.15$ to $B-V \approx 1.1$.  Some of the Landolt 
fields were observed several times during the night at different airmass in 
order to measure the atmospheric extinction coefficients. For the 
flatfielding we used skyflats exposed to about half the dynamic range of the 
CCD, and in general each flatfield used in the reductions was constructed as 
an average of about 5 individual integrations, slightly offset with respect
to each other in order to eliminate any stars.  

\section{Data reduction}
\label{sec:red}

  The initial reductions of all CCD images (bias subtraction, flatfield 
correction, removal of bad columns) were done using the standard tools in 
IRAF\footnote{
IRAF is distributed by the National Optical Astronomical Observatories,
which are operated by the Association of Universities for Research in
Astronomy, Inc. under contract with the National Science Foundation.
}. 

\subsection{Standard calibrations}

  The instrumental magnitudes for the standard stars were obtained on the 
CCD images using the {\bf phot} task in DAOPHOT (Stetson \cite{stet1987}) 
with an aperture radius of 20 pixels.
The extinction correction was then applied directly to the raw 
instrumental magnitudes.  The standard calibrations were carried out using the 
{\bf photcal} package in IRAF with the transformation equations assumed to 
be of the linear form

\begin{eqnarray*}
  V   & = & v + c_v \times (b-v) + z_v  \\
  B-V & = & c_{(b-v)} \times (b-v) + z_{(b-v)} \\
  U-B & = & c_{(u-b)} \times (u-b) + z_{(u-b)} \\
  V-R & = & c_{(v-r)} \times (v-r) + z_{(v-r)} \\
  V-I & = & c_{(v-i)} \times (v-i) + z_{(v-i)} \\
\end{eqnarray*}

  Here the magnitudes written in capital letters correspond to the standard
system, and magnitudes written with small letters are in the instrumental
system, corrected for extinction only. 

\begin{table}
\caption{
  The standard transformation coefficients for each of the four observing
runs. $N$ is the total number of observations of individual stars used
for the transformations. The number of stars $N_s$ actually involved is 
always smaller, some of the fields being observed several times. The rms
values refer to the scatter around the standard transformations.
  \label{tab:stdtrans}
}
\begin{tabular}{lrrrr} \hline
            &  NOT1    &  NOT2    &    DK1   &    DK2   \\
$N_s$       &   14     &   32     &     45   &     35   \\
$N$         &   25     &   69     &     98   &     94   \\
$c_v$       & $-0.056$ & $-0.034$ &   0.039  &   0.029  \\
$c_{(b-v)}$ &   1.109  &   1.111  &   1.100  &   1.104  \\
$c_{(u-b)}$ &   1.081  &   1.055  &   0.943  &   0.959  \\
$c_{(v-r)}$ &   1.012  &   1.027  &   1.013  &   1.014  \\
$c_{(v-i)}$ &   0.919  &   0.948  &   1.031  &   1.045  \\
$z_v$       & $-0.699$ & $-0.481$ & $-2.064$ & $-2.094$ \\
$z_{(b-v)}$ &   0.082  &   0.005  & $-0.540$ & $-0.432$ \\
$z_{(u-b)}$ & $-1.988$ & $-1.830$ & $-2.118$ & $-2.260$ \\
$z_{(v-r)}$ & $-0.091$ & $-0.086$ & $-0.019$ &   0.154  \\
$z_{(v-i)}$ &   0.219  &   0.234  &   1.065  &   1.198  \\ 
rms($V$)    &   0.008  &   0.034  &   0.019  &   0.037  \\
rms($B-V$)  &   0.013  &   0.042  &   0.019  &   0.032  \\
rms($U-B$)  &   0.042  &   0.043  &   0.050  &   0.041  \\
rms($V-I$)  &   0.017  &   0.033  &   0.030  &   0.024  \\ \hline
\end{tabular}
\end{table}

  The coefficients of the standard transformation equations are given in
Table~\ref{tab:stdtrans} for each of the four observing runs. NOT1 and
NOT2 refer to observing runs at the Nordic Optical Telescope in May 1997
and October 1997, respectively, and DK1 and DK2 refer to runs at the
Danish 1.54m telescope in September 1997 and February 1998. The
CCD at the NOT was interchanged between the two runs which may explain
the changes in the instrumental system from one run to the other, while the 
quantum efficiency of the CCD at the 1.54m telescope was notoriously unstable 
during the summer 1997/1998, and therefore the changes in the zero-point 
constants are hardly surprising. The rms scatter of the standard star
observations relative to the standard system after the transformation is 
generally of the order of a few times 0.01 mag., with a slightly larger 
scatter in the U-B transformation. Thus, the accuracy of the standard
transformations is sufficient for our purpose.

\subsection{Reduction of science data}

  Following the initial reductions, the three exposures in each filter 
were aligned and combined into a single image, yielding an effective 
integration time of 15 minutes in BVRI and H$\alpha$ and 60 minutes in U. 
As both the DFOSC and ALFOSC instruments can be rotated around the optical 
axis it was sometimes necessary to counter-rotate the individual exposures 
before they were combined (using the IRAF task {\bf rotate}), even if 
attempts were made to position the instruments in the same way from night 
to night. This is due to the limited angular resolution of the encoders on 
the rotators. The necessary offsets and rotations were calculated by 
manually inspecting each frame and measuring the x-y coordinates of two 
selected stars using the IRAF task {\bf imexamine}. By this method it was 
possible to align the images to a precision of about 0.1 pixel, accurate 
enough for our purpose. Flux conservation during the rotation was checked
by comparing photometry on a synthetically generated image and on a rotated
version of the same image. No deviations larger than 0.003 mag. were found.

  The actual combination of the individual images was done using the task 
{\bf imcombine}, with the {\bf reject} option set to {\bf crreject} in
order to remove cosmic ray events.
It was necessary to be particularly careful if the images to be combined 
had slightly different seeing, because the centres of bright stars could 
then be ``rejected'' by {\bf imcombine}. We checked that this did not happen 
by also combining the images using no rejection at all and subtracting the two 
versions of the combined images from each other, making sure that no 
residuals were seen at the positions of stars. In general it was possible to 
avoid this problem by adjusting the {\bf snoise} (``sensitivity noise'') 
parameter of {\bf imcombine}.

  As the first step towards detecting point-like objects we then created a 
smoothed version of a V-band image, using the
median filtering task in IRAF and a 5 $\times$ 5 pixels box median filter.
The median-filtered image was then subtracted from the original version,
resulting in a residual image that contained no large-scale structure but only
point sources superimposed on a uniform background.  Point sources were
finally located in the residual image using the {\bf daofind} task in DAOPHOT.

  Because star clusters are sometimes seen as slightly extended sources 
even at the distance of the galaxies in our sample, the photometry had
to be obtained as aperture photometry rather than by PSF fitting (see
also {\"O}stlin et al. \cite{ostlin1998}). The photometry was carried out 
directly on the combined images, and not on background subtracted images.
  In Paper1 it was argued that the accuracy obtained by this method was quite 
satisfactory.  
An aperture radius of 4 pixels was chosen for the determination 
of colour indices, while an aperture radius of 8 pixels was used to measure 
$V$ band magnitudes.  It was shown in Paper1 (see also e.g. Holtzman et al.
\cite{hol1996}) that aperture corrections tend to cancel out for colour 
indices, thus making it possible to measure accurate colours even with a 
quite small aperture.

The pixel size on the NOT images is half of that on the 
1.54 m. images, but by coincidence the seeing on the NOT data was also, in
general, twice as good as on the 1.54 m. data, so we used the same aperture 
radii in pixels for the NOT and 1.54 m. observations.  Aperture corrections 
relative to the 20-pixels aperture radius used for the standard star 
observations were applied to the photometry, based on the assumption that 
the objects were point sources.  The aperture corrections were derived
from photometry of a few bright, isolated stars using the {\bf mkapfile} 
task in DAOPHOT, and amounted in general to a few tenths of a magnitude. 
The photometry was corrected for Galactic foreground reddening using
the extinction values given in the RC3 catalogue (de Vaucouleurs et al. 
\cite{vau91}).

\subsection{Identification of star clusters}
\label{subsec:identi}

\begin{figure}
\epsfxsize=8.8cm
\epsfbox[85 370 552 720]{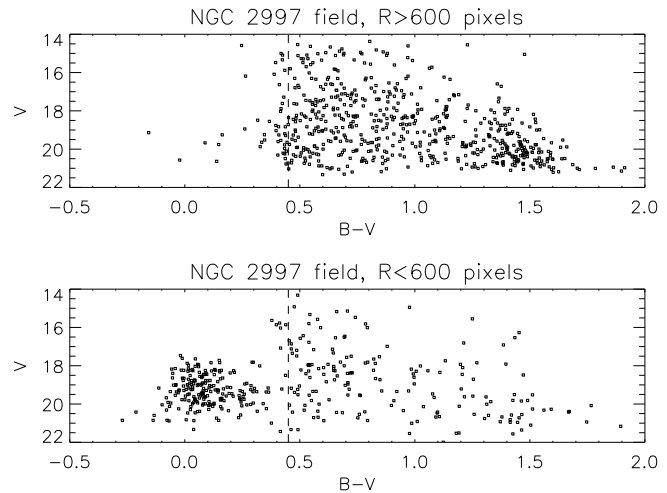}
\caption{
  Colour-magnitude diagrams for objects in the NGC~2997 field. Top: Objects
at distance larger than 600 pixels (4 arcminutes) from the centre of the 
galaxy. Bottom: Objects at distance closer than 600 pixels to the centre 
of the galaxy.
\label{fig:bvcrit}
}
\end{figure}

  In short, young star cluster candidates were identified as compact bright, 
blue objects without any $H\alpha$ emission. The specific criteriae
invoked to select cluster candidates were the following:

  The first criterion was that the objects should have \mbox{$B-V < 0.45$}.
In this way most foreground stars are eliminated, as demonstrated
in Fig.~\ref{fig:bvcrit} which shows colour-magnitude diagrams for objects
in the field containing the galaxy NGC~2997. The NGC~2997 field is 
relatively rich in foreground stars, while the galaxy itself occupies only 
the central part of the frame, and therefore this field serves as an 
illustrative example.  The difference between the colour-magnitude diagrams 
of the central part of the field (where the galaxy is located) and the outer
parts (only foreground) is striking. In fact some of the foreground stars
are slightly bluer than \mbox{$B-V = 0.45$}, so the limit is not completely
rigorous and was adjusted slightly from case to case by inspecting 
diagrams like the one in Fig.~\ref{fig:bvcrit}. A \mbox{$B-V$} colour of 
0.45 for a star cluster corresponds to an age of about 500 Myr 
(Girardi et al. \cite{girardi1995}), so we are really just sampling young 
objects.

  Secondly, the cluster candidates were divided into two groups, a 
``red'' group with $U-B > -0.4$ and a ``blue'' group with $U-B \le -0.4$,
and an absolute magnitude limit was applied.  For a given mass, clusters 
in the ``blue'' group will be more luminous than clusters in the ``red''
group, and therefore a brighter magnitude cut-off could be applied to
the ``blue'' clusters. Specifically, the limits were set to
$M_V < -8.5$ for the ``red'' group and to $M_V < -9.5$ for the ``blue''
group (distance modulii for the galaxies are given in Paper1).

\begin{figure}
\epsfxsize=88mm
\epsfbox[90 370 550 720]{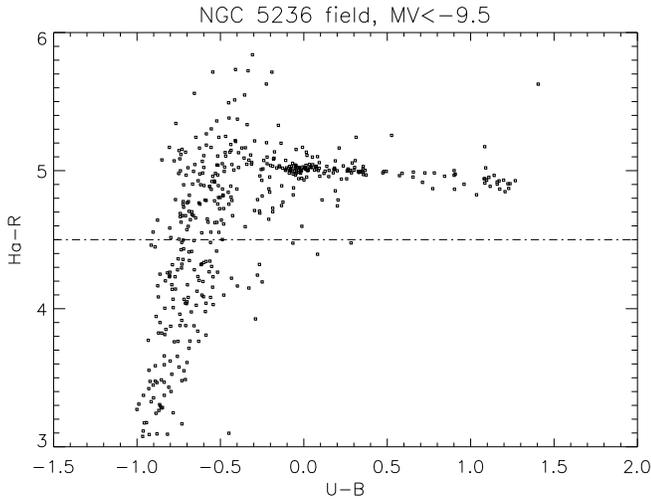}
\caption{
  $H\alpha-R$ vs. $U-B$ for objects brighter than $M_V = -9.5$ in the
NGC~5236 field.
\label{fig:haplot}
}
\end{figure}

  Third, it was necessary to exclude HII regions.  In Fig.~\ref{fig:haplot} 
we show $H\alpha-R$ vs. \mbox{$U-B$} for all objects brighter than 
$M_V = -9.5$ in the NGC~5236 field, chosen as an example because of the 
large number of sources in this galaxy. The $H\alpha$ photometry was not 
standard calibrated, so the magnitude scale on the y-axis in 
Fig.~\ref{fig:haplot} is purely instrumental. It is clear from 
Fig.~\ref{fig:haplot} that the objects with an excess in $H\alpha$ appear 
around $U-B = -0.4$, while $H\alpha-R$ is rather constant for redder 
objects. Denoting the average $H\alpha-R$ value at $U-B > -0.4$ by 
$(H\alpha-R)_{\mbox{\scriptsize ref}}$, we chose to cut away objects with
$H\alpha-R < (H\alpha-R)_{\mbox{\scriptsize ref}} - 0.5$.
The limit could have been placed closer to 
$(H\alpha-R)_{\mbox{\scriptsize ref}}$, but by choosing a cut-off at 0.5 
magnitudes below $(H\alpha-R)_{\mbox{\scriptsize ref}}$ we allow for some 
scatter in the photometry as well.

  Finally we did a visual inspection of the cluster candidates. In this 
way objects which were too extended or ``fuzzy'' to be star clusters were 
eliminated (typically star clouds or associations in the spiral arms of 
the galaxies).

\begin{figure}
\begin{minipage}[t]{17mm}
n1313-Psf\\
\epsfxsize=17mm
\epsfbox{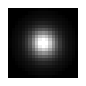}\\
n5236-Psf\\
\epsfxsize=17mm
\epsfbox{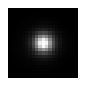}\\
n628-Psf\\
\epsfxsize=17mm
\epsfbox{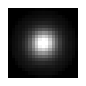}\\
\end{minipage}
\begin{minipage}[t]{17mm}
n1313-201\\
\epsfxsize=17mm
\epsfbox{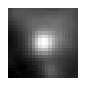}\\
n5236-76\\
\epsfxsize=17mm
\epsfbox{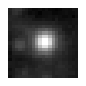}\\
n628-274\\
\epsfxsize=17mm
\epsfbox{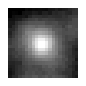}\\
\end{minipage}
\begin{minipage}[t]{17mm}
n1313-379\\
\epsfxsize=17mm
\epsfbox{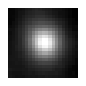}\\
n5236-175\\
\epsfxsize=17mm
\epsfbox{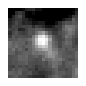}\\
n628-1171\\
\epsfxsize=17mm
\epsfbox{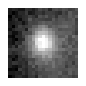}\\
\end{minipage}
\begin{minipage}[t]{17mm}
n1313-574\\
\epsfxsize=17mm
\epsfbox{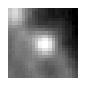}\\
n5236-199\\
\epsfxsize=17mm
\epsfbox{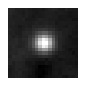}\\
n628-1880\\
\epsfxsize=17mm
\epsfbox{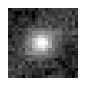}\\
\end{minipage}
\begin{minipage}[t]{17mm}
n1313-627\\
\epsfxsize=17mm
\epsfbox{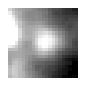}\\
n5236-347\\
\epsfxsize=17mm
\epsfbox{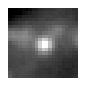}\\
n628-2283\\
\epsfxsize=17mm
\epsfbox{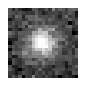}\\
\end{minipage}
\caption{\label{fig:clfig} V-band images of a few clusters. The first image in
each row is the PSF, followed by 4 clusters.}
\end{figure}

  Photometric data are given for each cluster in Table~\ref{tab:clusters}, 
along with the effective radius $R_e$ (in pc) as derived by {\bf ishape} 
(Sec.~\ref{sec:tools}). A few examples of clusters are shown in
Fig.~\ref{fig:clfig}.

  A table showing the total number of star clusters identified in each
of the galaxies in our sample was given in Paper1, and
we refer to that paper for a detailed discussion of the properties of
the cluster systems. 

\section{Tools for synthetic image generation and analysis}
\label{sec:tools}

  It is a complicated problem to carry out photometry on star clusters 
located within spiral galaxies, partly because of the strongly varying
background, and partly because the clusters are not perfect point sources.
The measurements are subject to many potential errors, and it is
essential to check the photometry carefully and get a
realistic idea of the achievable accuracy. One way to do this is to
carry out experiments with artificial objects, which can be
added at any desired position in the image and remeasured using the
photometric method of choice. In order to give realistic estimates of
the photometric errors the artificial objects should, of course, resemble
the real objects as closely as possible.

  We have developed a number of tools to be used in the analysis of this
type of image data. The two most important ones, which will be described 
below are 

\begin{itemize}
  \item {\bf mksynth} - generates a synthetic image by adding artificial 
          sources to a uniformly illuminated sky background. 

  \item {\bf ishape} - which finds the
          intrinsic shape of extended sources by modeling them as
          one of several available analytic models.
          The parameters of the analytic model are adjusted until 
	  the best fit between the observed profile and the model convolved 
	  with the PSF of the image is obtained.
\end{itemize}

  In practice these two algorithms (together with some more general image
processing functions) are built into one stand-alone programme\footnote{
The programme containing the routines, (for historical reasons called 
{\bf baolab}), is available at the
following WWW address: http://www.astro.ku.dk/$^\sim$soeren/baolab/
}, so that 
they can share common routines to handle user-definable parameters, read 
and write FITS images etc. 

  We have not included a task to generate the PSF itself from an image. 
This must be done using some other programme, such as the {\bf psf} and
{\bf seepsf} tasks in DAOPHOT. We refer to the IRAF and DAOPHOT documentation 
for more details on these topics.

\subsection{mksynth}

  This algorithm generates synthetic images which are as similar to real 
data as possible, including a ``sky background'' with gaussian photon shot 
noise.  Stars are generated not just by adding a scaled PSF but rather by 
a process similar to that by which the photons arrive in a real CCD image 
during an integration. Thus, in contrast to other popular algorithms for 
adding synthetic stars to an image (such as {\bf addstar} in DAOPHOT), 
{\bf mksynth} generates a complete synthetic image from scratch, including 
a noisy background if desired. The PSF can be modeled as one of several 
analytic profiles (see Sect.~\ref{subsec:ishape}), or read from a FITS file. 

  One of the major forces of the algorithm is that it allows a great
flexibility in the generation of synthetic images, through a number of
user-definable parameters. The coordinates and magnitudes of the 
synthetic objects can be read from a file, or generated at random.

  Tests have shown that synthetic images generated by {\bf mksynth}
have very realistic noise characteristics, and it is possible to generate
synthetic images which resemble real CCD images very closely. {\bf mksynth}
was described and tested more fully (although in a more primitive version)
by Larsen (\cite{larsen1996}).

\subsection{ishape}

\label{subsec:ishape}

\begin{table}
\caption{Essential {\bf ishape} parameters.
  \label{tab:ishapepar}
}
\begin{tabular}{lp{43mm}}\hline
FITRAD = float & Fitting radius (in pixels) \\
CENTERRAD = float & Centering radius \\
CLEANRAD = float & Cleaning radius \\
CTRESH = float & Cleaning threshold (in $\sigma$) \\
SHAPE = string & Which model to use \\
ELLIPTICAL = YES/NO & Fit elliptical model? \\
EPADU = float & Conversion factor of CCD image \\
RON = float & Read-out noise \\
RESIMAGE = string & File with residual image \\
REPLACE = YES/NO & Remove the object from the original image? \\
KEEPLOG = YES/NO & Keep a log file? \\
LOGFILE = string & Name of log file (if KEEPLOG=YES) \\
\hline
\end{tabular}
\end{table}

  {\bf ishape} can be used to estimate the
intrinsic shape parameters of extended objects in a digital image with a 
known PSF. The algorithm is designed to work in the domain of ``slightly'' 
extended objects which can be modeled as simple analytic functions, i.e. 
objects with a size roughly equal to or smaller than the PSF. Conventional
deconvolution algorithms are not designed for this type of problem. None
of the ``first-generation'' deconvolution algorithms such as the Maximum
Entropy Principle (Burch et al. \cite{burch1983}) and the Richardson-Lucy 
algorithm (Richardson \cite{richard1972}, Lucy \cite{lucy1974}) handled 
point-like sources well at all. More recent ``two-channel algorithms'' 
(Lucy \cite{lucy1994}, Magain et al. \cite{magain1998}) 
model the image as consisting of a smoothly varying background and a number 
of $\delta$-functions. The two-channel algorithms
seem to work quite well in many cases, being able to separate point sources 
and obtain deconvolved images of photometric quality, but they are not able 
to treat objects which are only {\it nearly} point-like. Therefore we feel 
it is worthwhile to spend some space describing our algorithm which handles
this specialised, but for our work important case. We have used {\bf ishape}
to derive intrinsic radii for star clusters in other galaxies, but one
could also imagine other areas of work where the algorithm might be useful, 
for example in the study of distant galaxies which are just barely resolved.

  The analytic profiles by which {\bf ishape} models the sources are:

\begin{eqnarray}
 \mbox{GAUSS:} & S(z) = & \exp(-z^2) \\
 \mbox{MOFFAT15:} & S(z) = & \frac{1}{(1+z^2)^{1.5}} \\
 \mbox{MOFFAT25:} & S(z) = & \frac{1}{(1+z^2)^{2.5}} \\
 \mbox{KINGc:} & S(z) = & \left\{ \begin{array}{ll}
				 \left( \frac{1}{\sqrt{1+z^2}} 
				-\frac{1}{\sqrt{1+c^2}} \right)^2 & 
				  \mbox{for } z < c \\
                                  0 & \mbox{for } z \ge c
				  \end{array}
                          \right. \\
 \mbox{LUGGER:} & S(z) = & \frac{1}{(1+z^2)^{0.9}} \\
 \mbox{HUBBLE:} & S(z) = & \frac{1}{(1+z^2)} \\
 \mbox{DELTA:}  & S(z) = & \delta (z)
\end{eqnarray}

Here $z$ is given by
the equation $z^2 = a_1 x^2 + a_2 y^2 + a_3 x y$, where the constants
$a_1$, $a_2$ and $a_3$ depend on the major axis, ellipticity and
orientation of the model, and $x$ and $y$ are the coordinates relative
to the centre of the profile.  

  For the KING models, the concentration 
parameter\footnote{For convenience, we define the concentration 
parameter as $c = r_t/r_c$ where $r_t$ is the tidal radius and $r_c$ is the 
core radius. In the literature it is customary to define $c = \log(r_t/r_c)$.
}
$c$ may assume the 
values 5, 15, 30 and 100. Note that the HUBBLE model is equal to a KING model
with infinite concentration parameter. 
  The MOFFAT models are similar to the profiles used by Elson et al. 
(\cite{elson1987}) 
to fit young LMC clusters, with their $\gamma = 3$ profile corresponding
to the MOFFAT15 model and $\gamma = 5$ to the MOFFAT25 model. Elson et 
al. (1987) found $2.2 < \gamma < 3.2$ for their sample of LMC clusters. 
Unlike the KING models, the MOFFAT functions never reach a value of 0, but
both the MOFFAT15 and MOFFAT25 functions share the desirable property that 
their volume is finite so that a well-defined effective radius exists.
Clearly, the DELTA model is normally of little use, and is mostly used 
internally by {\bf ishape}. The code can easily be extended to include other 
models as long as they can be described as simple analytic functions of the 
parameter $z$.

  Denoting the observed image of an object $I$, the PSF with $P$, the 
intrinsic shape with $S$ and the convolution of the two $M = P \star S$,
the algorithm finds $S$ by minimising the function:
\begin{equation}
  \chi^2 = 
    \sum_{i,j}W_{ij}\left[(I_{ij} - 
              M_{ij}(x,y,a,b,w_x,w_y,\alpha))/\sigma_{ij}\right]^2
  \label{eq:chis2}
\end{equation}
  $x,y$ is the position of the object, $a$ and $b$ represent the
amplitude (brightness) and the background level, and $w_x$, $w_y$ and
$\alpha$ are the FWHM along the major and minor axes and the position
angle. $\sigma_{ij}$ is the statistical uncertainty on the pixel
value at the position $(i,j)$, and $W_{ij}$ is a weighting function. 
The summation (\ref{eq:chis2}) is carried 
out over all the elements $(i,j)$ of the image area considered (typically 
a rather small section around the object). 

  The actual implementation
of the minimisation of Eq.~(\ref{eq:chis2}) is somewhat more complicated
than just minimising $\chi^2$ as a function of all seven parameters at
once, and particular care is taken to evaluate the convolution
$M = P \star S$ as few times as possible. Only the parameters
$w_x, w_y$ and $\alpha$ affect the the actual shape of the convolved
profile, so one might think of $\chi^2$ as a function
of these three parameters only, with the minimisation of the remaining
parameters ($x,y$ position, the amplitude $a$ and background $b$) being
carried out implicitly for each choice of $w_x, w_y$ and $\alpha$. 
The function $\chi^2(w_x, w_y, \alpha)$ is minimised using the ``downhill 
simplex'' algorithm (Press et al. \cite{press}) which has the advantage of 
being simple and robust. The initial guesses are partly user-definable, but
tests have shown that as long as convergence is reached, the results
are insensitive to the initial guesses. 

  The result of the fit is given as a FWHM along the major axis and an
axis ratio and orientation, but the FWHM may easily be converted to an
effective radius (containing half the total cluster light) for all
profiles except the HUBBLE and LUGGER profiles.

  Both in {\bf mksynth} and {\bf ishape}, the arrays containing image data 
are in reality stored internally with a resolution 10 times higher than the 
actual image resolution.  However, when calculating the $\chi^2$ 
(Eq.~(\ref{eq:chis2})) the arrays are rebinned to the original resolution.

  The weighting array $W$ is introduced in order to reduce the effect of
bad pixels, cosmic ray events, nearby stars etc. The weights $W$ are
derived from the input image {\it before the iterations are started} by 
calculating the standard deviation among the pixels located in concentric 
rings around the centre of the object, and assigning a weight to each pixel 
which is inversely proportional to its deviation from the mean of the 
pixels located in the same ring.  If the deviation of $I_{ij}$ is smaller 
than one $\sigma$ then the corresponding weight $W_{ij}$ is set equal to 1, 
and if the deviation is larger than a user specified parameter (CTRESH) 
then $W_{ij}$ is set equal to 0, effectively rejecting that pixel. For small 
distances from the centre of the profile the statistics will become poor 
and hence all weights are set equal to 1
for distances smaller than a user-specified limit (CLEANRAD). It is clear 
that this method of assigning weights works best for images where the sources 
are more or less circularly symmetric - if this is not the case, then the 
CLEANRAD parameter should be set to a large value so that all pixels are
assigned equal weights.

\begin{figure}
\begin{minipage}{43mm}
\epsfxsize=43mm
\epsfbox{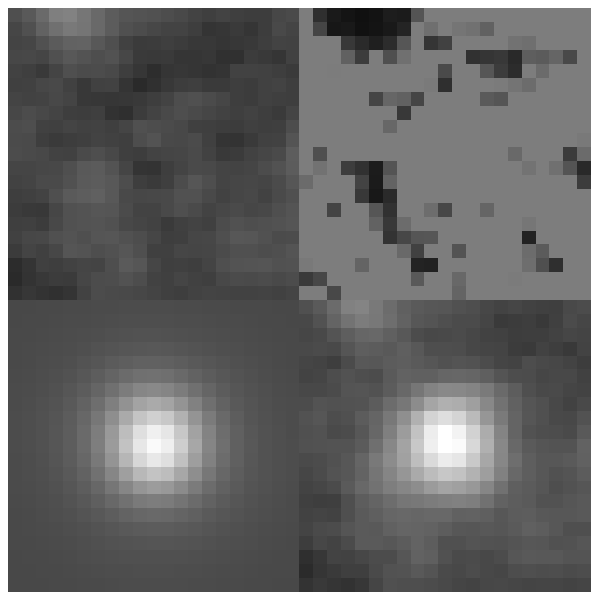}
\end{minipage}
\begin{minipage}{43mm}
\epsfxsize=43mm
\epsfbox{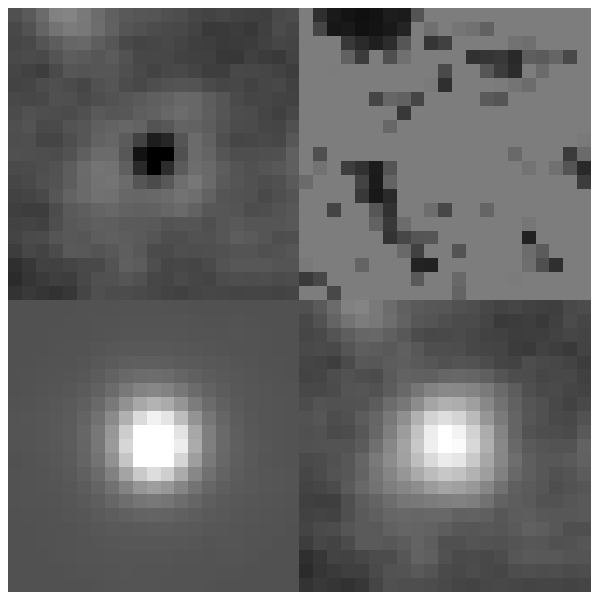}
\end{minipage}
\caption{
Residuals from {\bf ishape}, modeling a star cluster in NGC~5236. 
{\it Left:} The cluster was fitted using a MOFFAT15 model. {\it Right:} 
The cluster was fitted using a DELTA model.
\label{fig:ishape_res}
}
\end{figure}

  Examples of the output produced by {\bf ishape} are shown in 
Fig.~\ref{fig:ishape_res}. In the left part of the figure the cluster
was modeled as a MOFFAT15 function, and in the right part of the figure
the cluster was modeled as a DELTA function. In each set of four images
the original cluster $I$ is shown in the lower right corner, the final model
convolved by the PSF ($M$) is seen to the lower left, the fit residuals are
given to the upper left, and the weighting array $W$ is shown to the upper
right. Note how structures in the background correspond to regions that
are assigned a low weight, indicated by dark areas in the weighting
array.  In a typical situation it would have been adequate to choose
a smaller fitting radius (and thereby reduce the computation time), but
in this example we have extended the fitting radius to 11 pixels in order
to demonstrate how the star near the upper left corner affects the weighting
array. 

  From Fig.~\ref{fig:ishape_res} we note two things: First, the fit is
improved enormously by allowing the model to be extended as opposed to
the DELTA model, which corresponds to subtraction of a pure PSF. Hence,
the object is clearly recognised as an extended source. Second, considering
that the fitting radius in this example is as large as 11 pixels, the
residuals resulting from modeling the cluster as an extended source show
no other systematic variations than what can be attributed to background 
variations. In this particular example the
FWHM along the major axis of the MOFFAT15 function was found to be
1.67 pixels, and the FWHM of the PSF was 4.1 pixels.

\subsection{Tests of {\bf ishape}}
\label{subsec:itest}

  Before applying {\bf ishape} to data, it is of interest to know how 
reliably the elongation and major axis of small objects in a CCD image
can be reconstructed.  

\begin{figure} %
\begin{minipage}{43mm}
\epsfxsize=43mm
\epsfbox{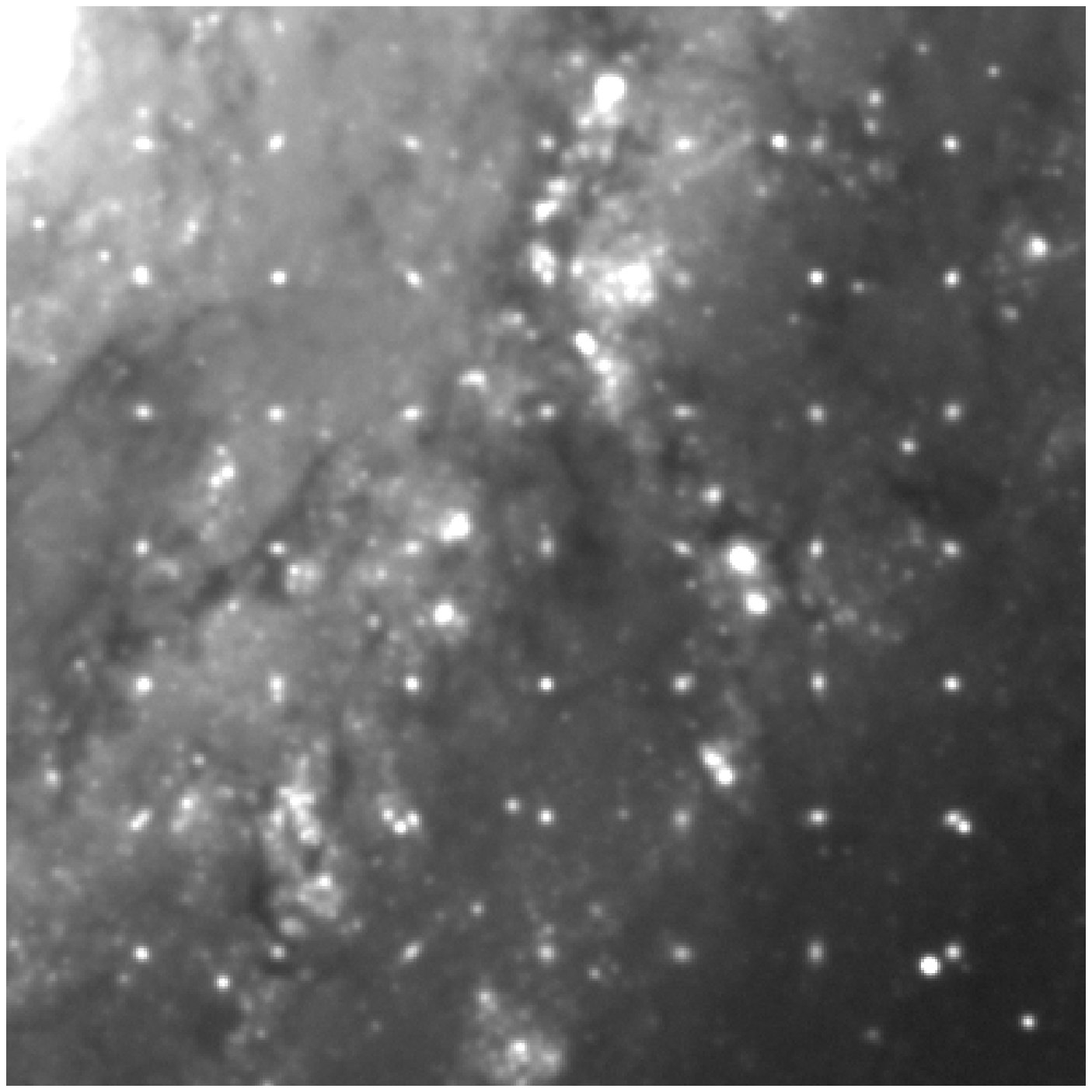}
\end{minipage}
\begin{minipage}{43mm}
\epsfxsize=43mm
\epsfbox{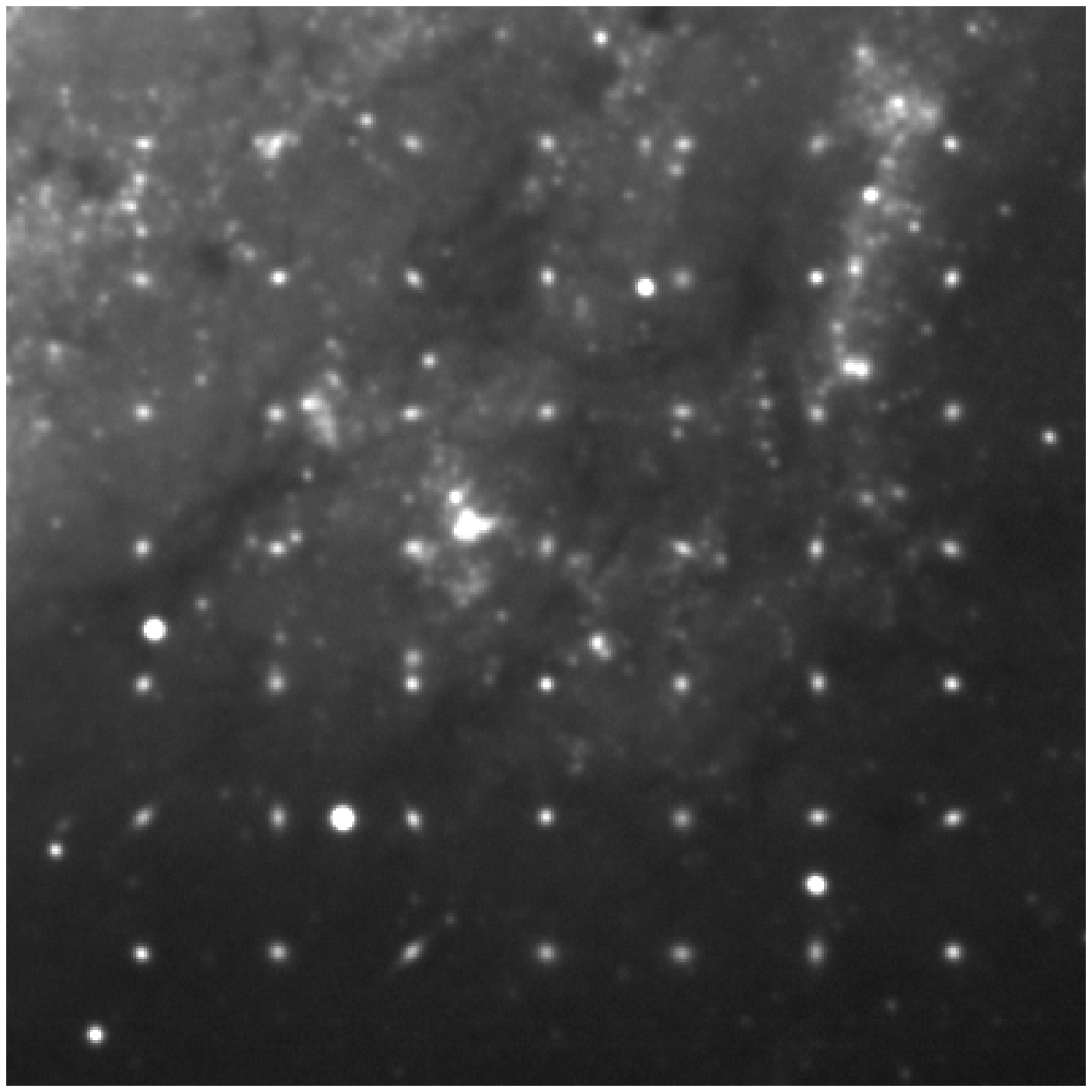}
\end{minipage}
\caption{
  The images of NGC~5236 used in the tests of {\bf ishape}.
{\it Left:} Inner field. {\it Right:} Outer field.  In this figure only the
images with the $m=18.7$ objects are shown. Note that most of the artificial
objects are too compact for their extent to be immediately visible.
\label{fig:testfields}
}
\end{figure}

\begin{figure}
\begin{minipage}{43mm}
\epsfxsize=43mm
\epsfbox[110 375 530 715]{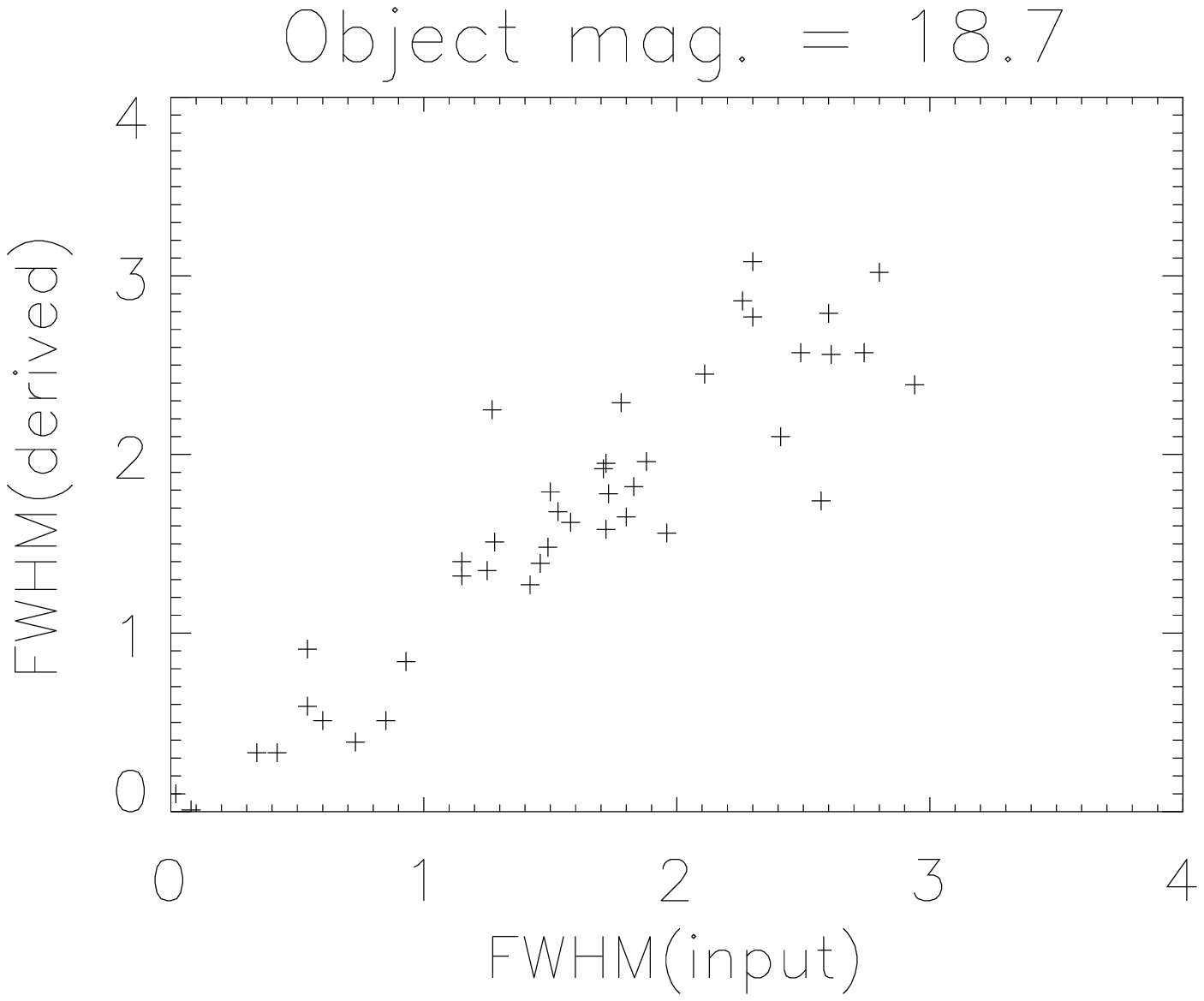}
\epsfxsize=43mm
\epsfbox[110 375 530 715]{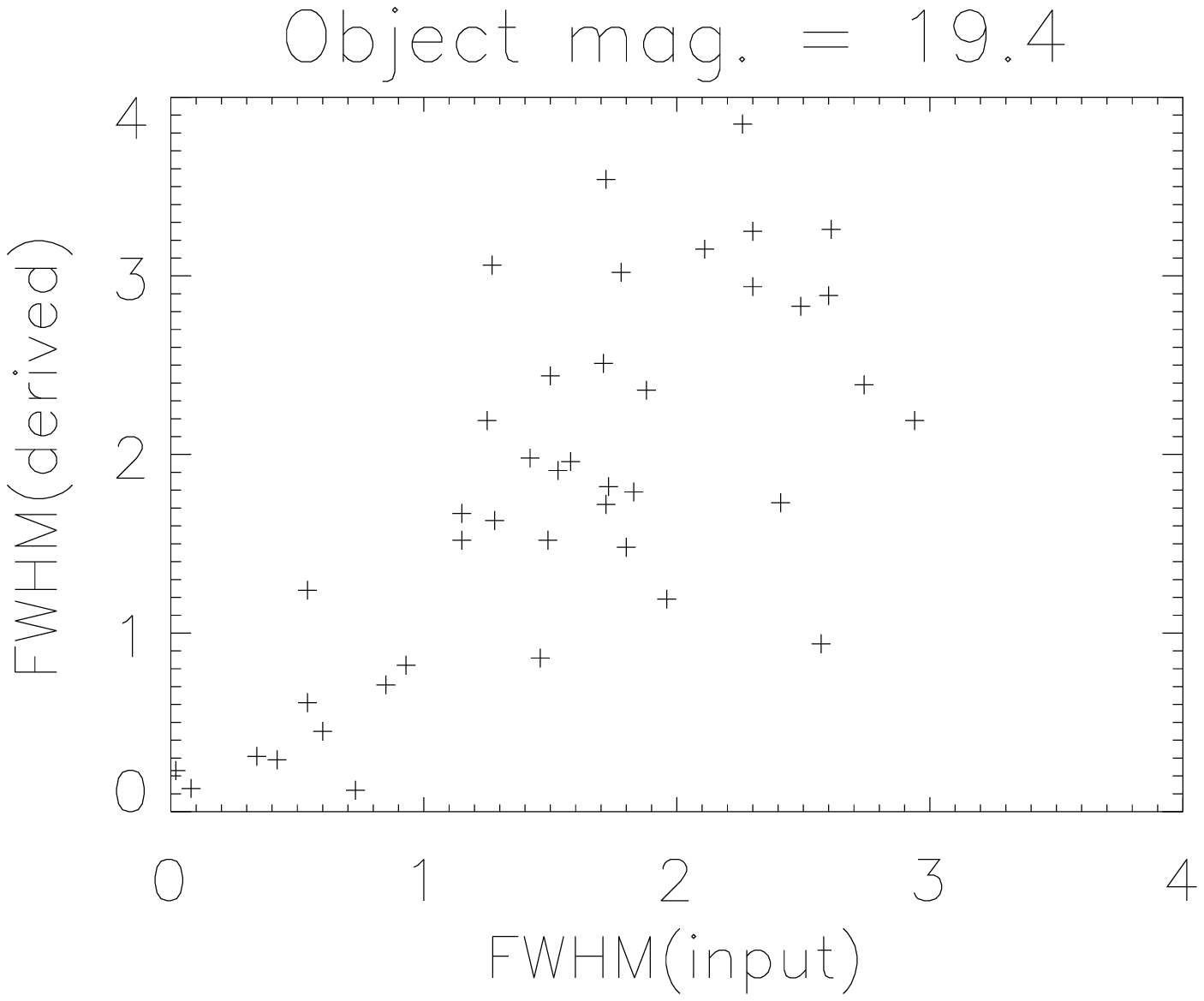}
\epsfxsize=43mm
\epsfbox[110 375 530 715]{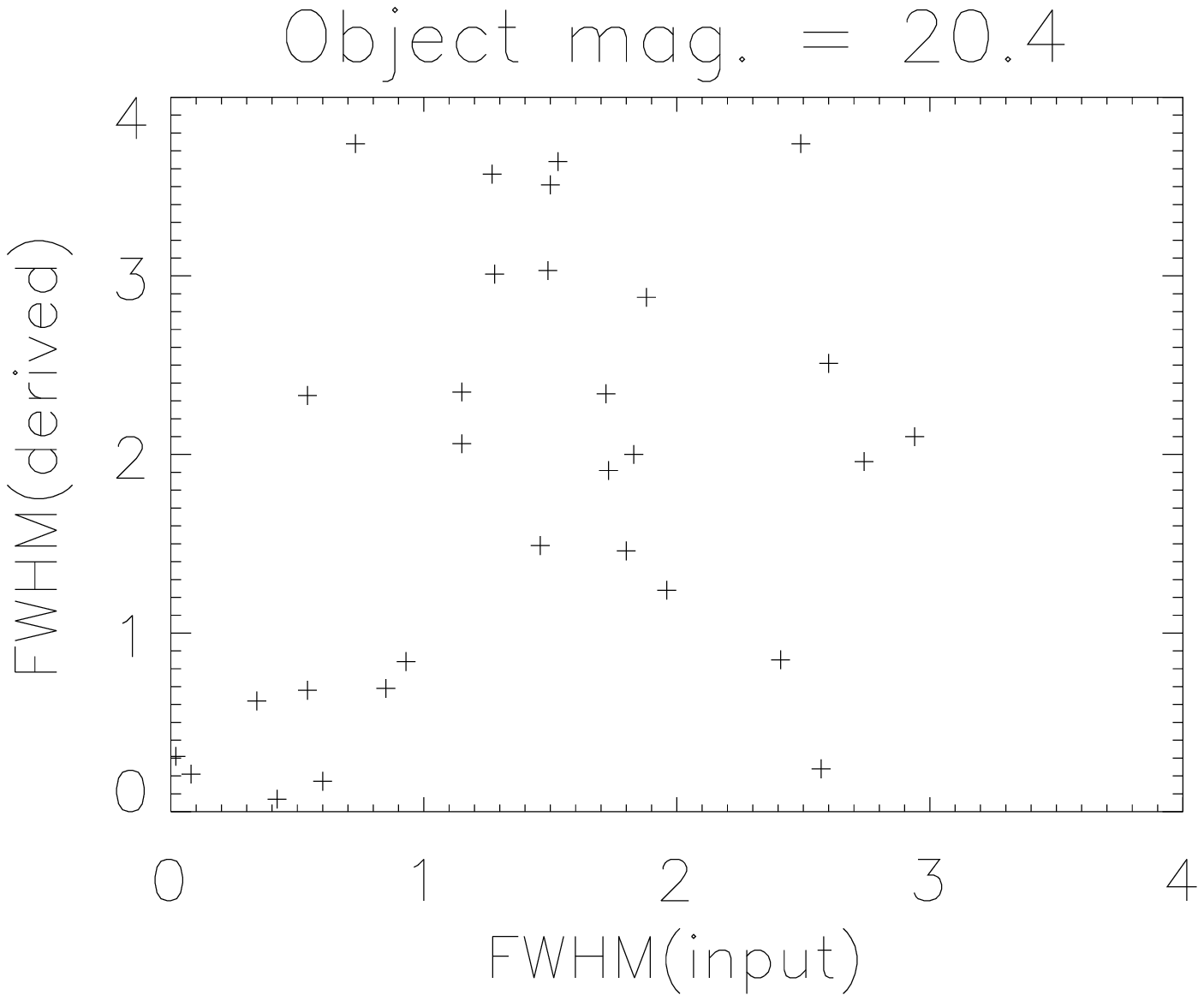}
\end{minipage}
\begin{minipage}{43mm}
\epsfxsize=43mm
\epsfbox[110 375 530 715]{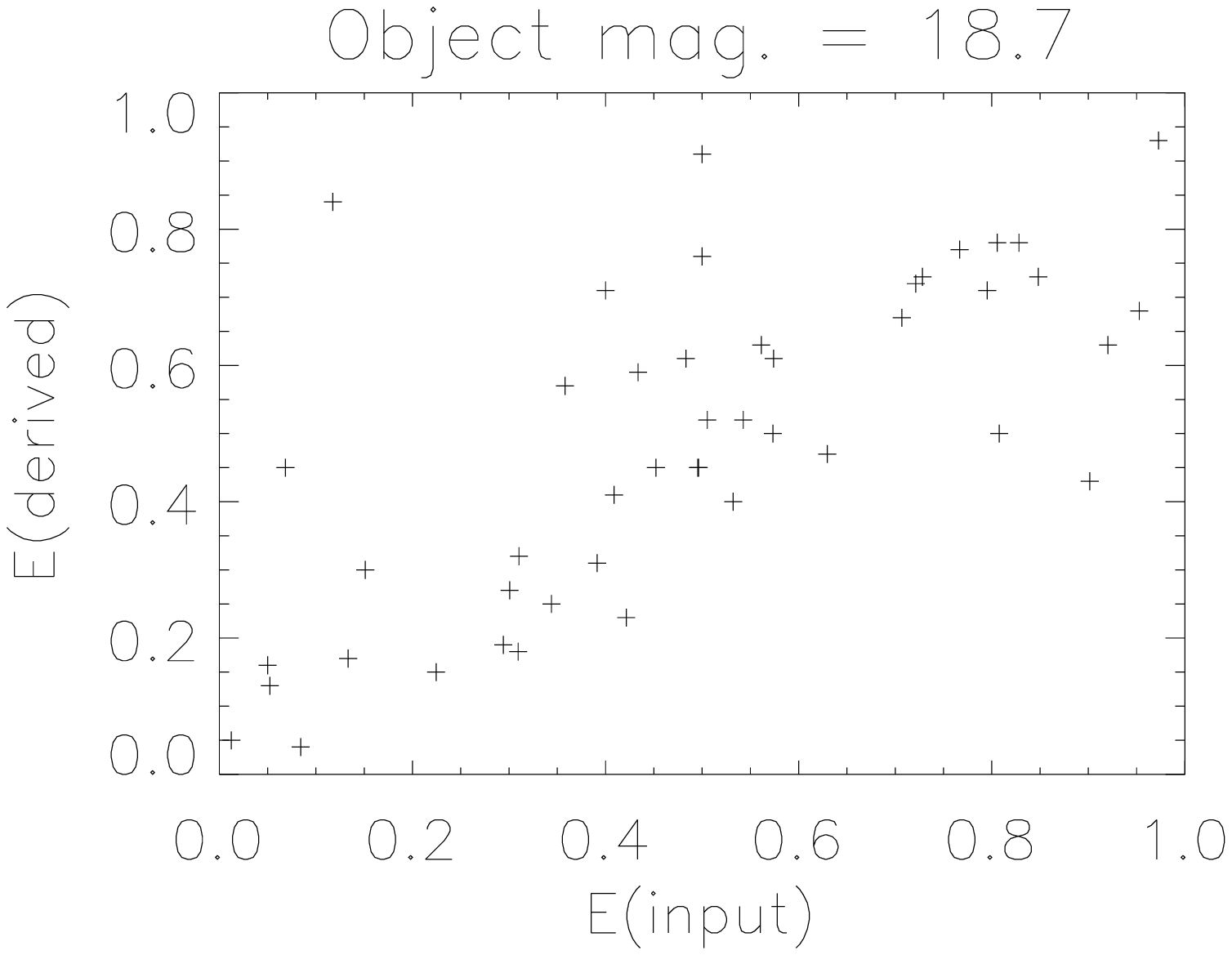}
\epsfxsize=43mm
\epsfbox[110 375 530 715]{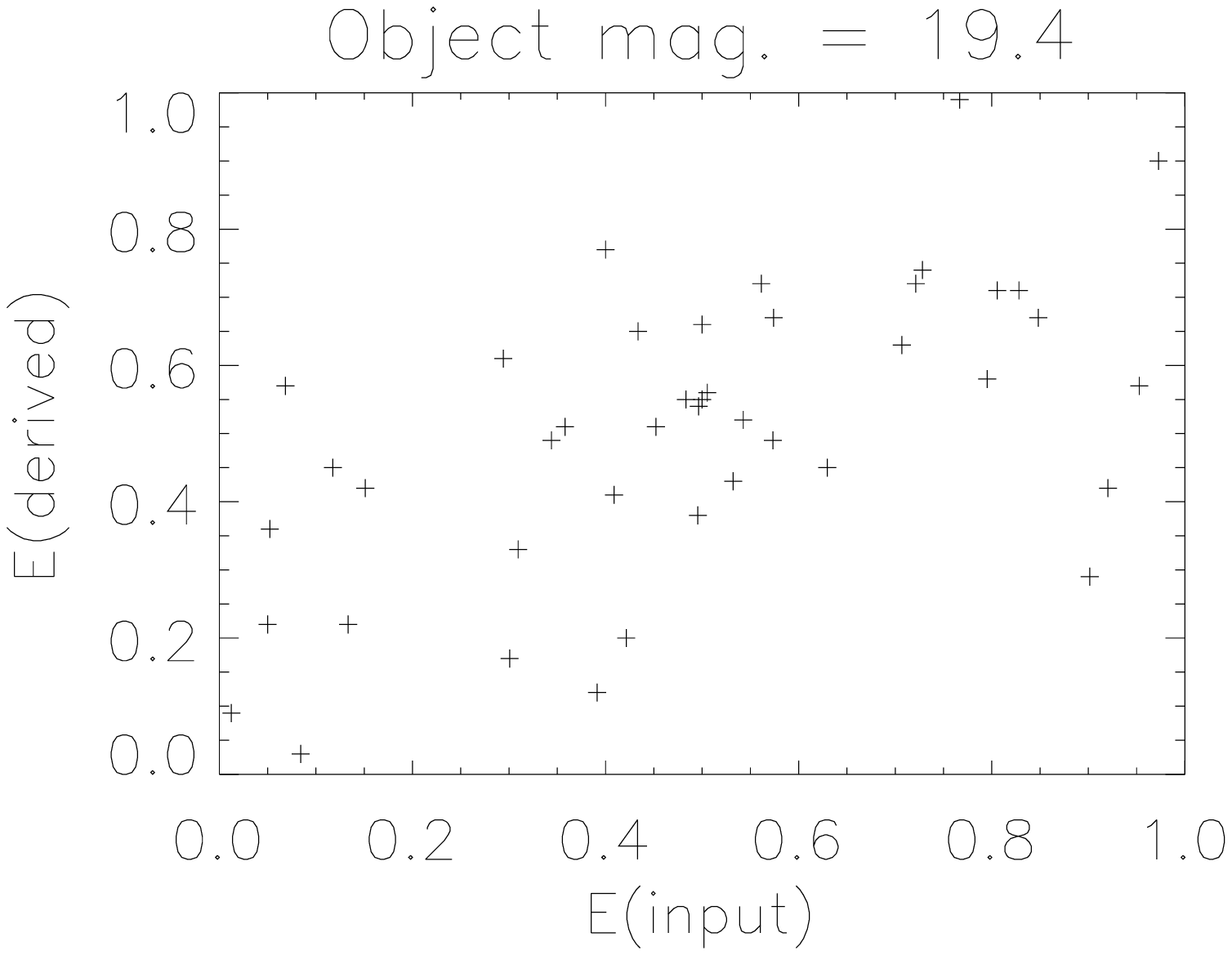}
\epsfxsize=43mm
\epsfbox[110 375 530 715]{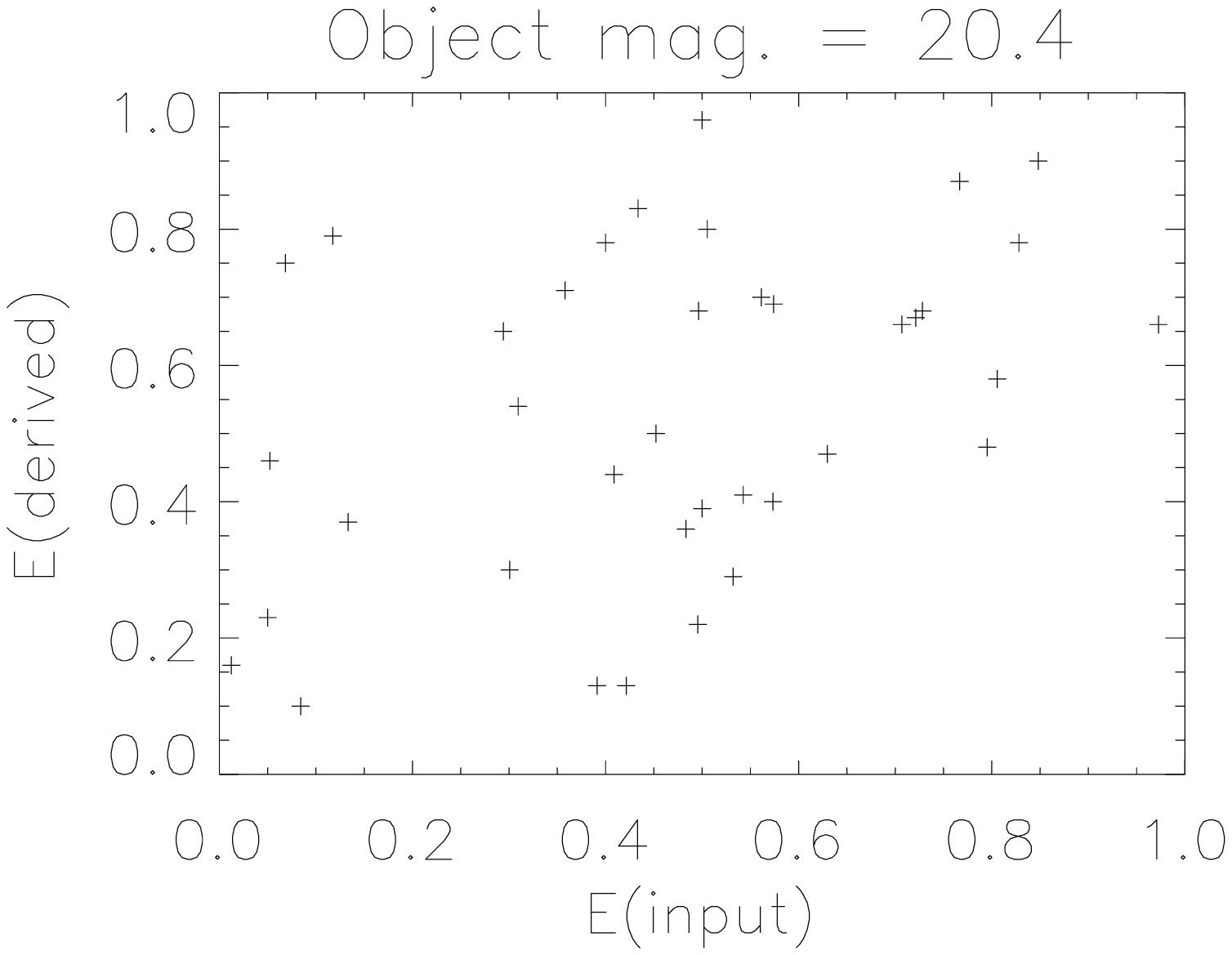}
\end{minipage}
\caption{
  Tests of {\bf ishape} using synthetic objects added to the \object{NGC~5236}
inner field. Input FWHM values (left) and axis ratios (right) are
compared to the values measured by {\bf ishape}.
  \label{fig:ishape1}
}
\end{figure}

\begin{figure}
\begin{minipage}{43mm}
\epsfxsize=43mm
\epsfbox[110 375 530 715]{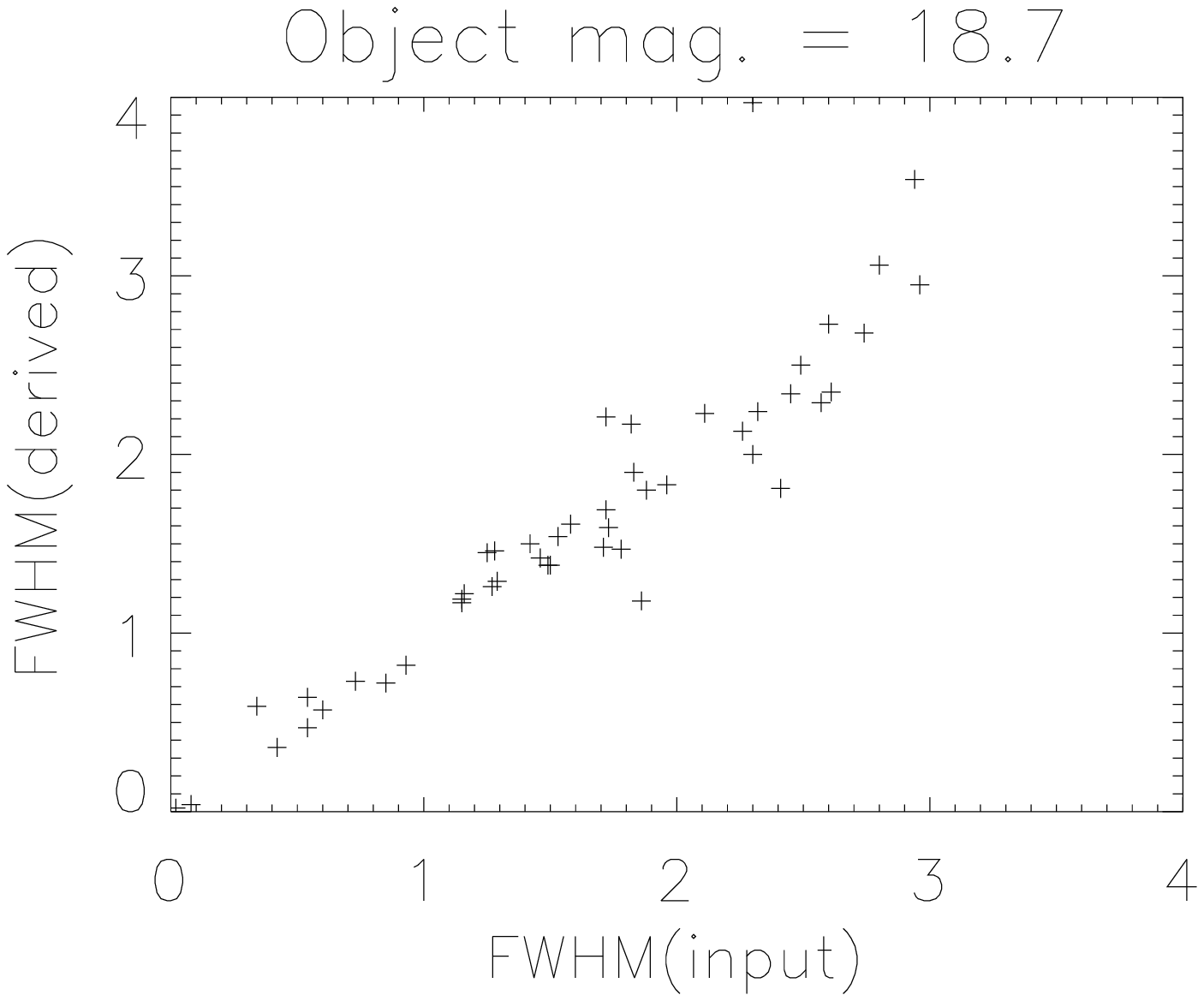}
\epsfxsize=43mm
\epsfbox[110 375 530 715]{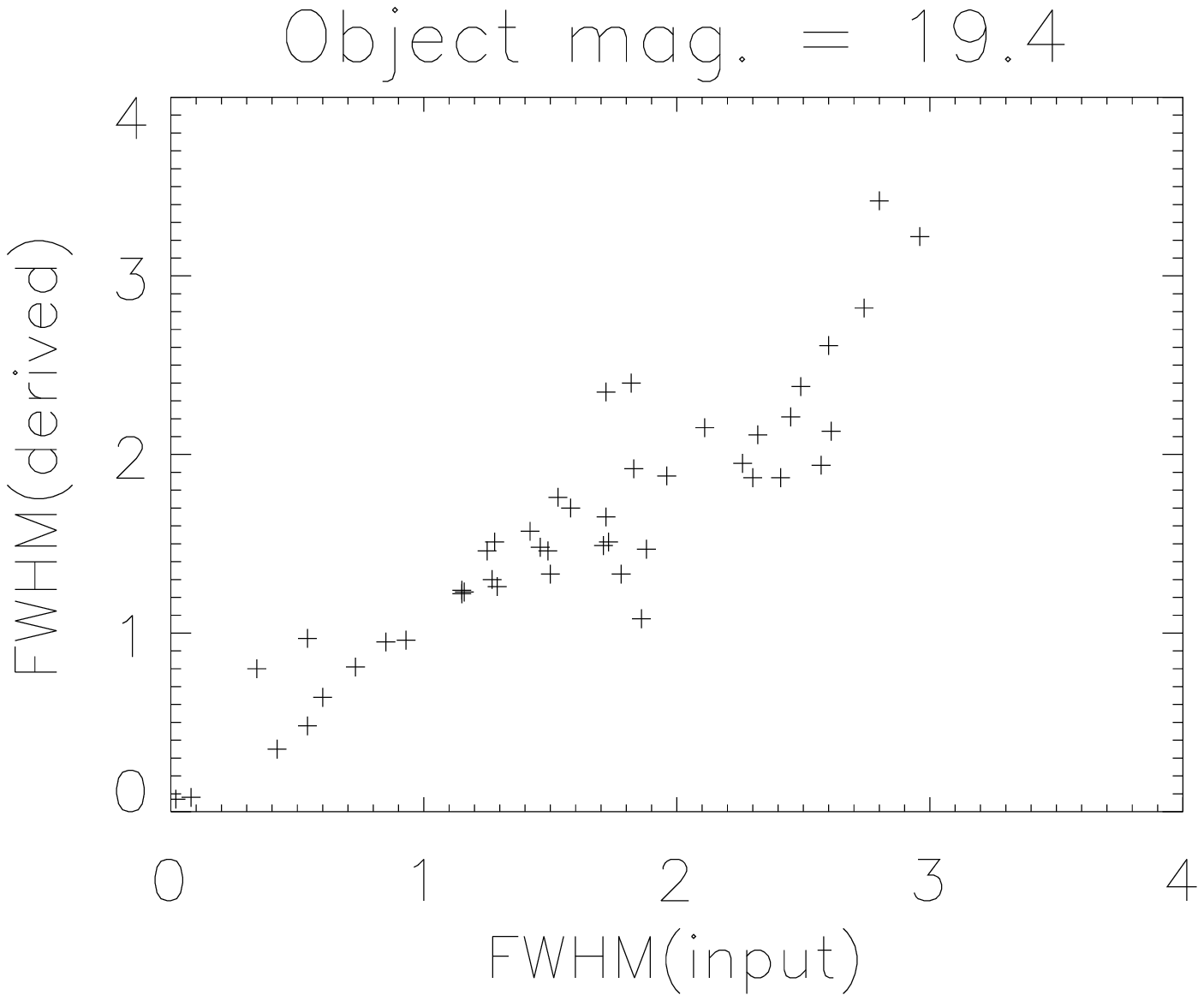}
\epsfxsize=43mm
\epsfbox[110 375 530 715]{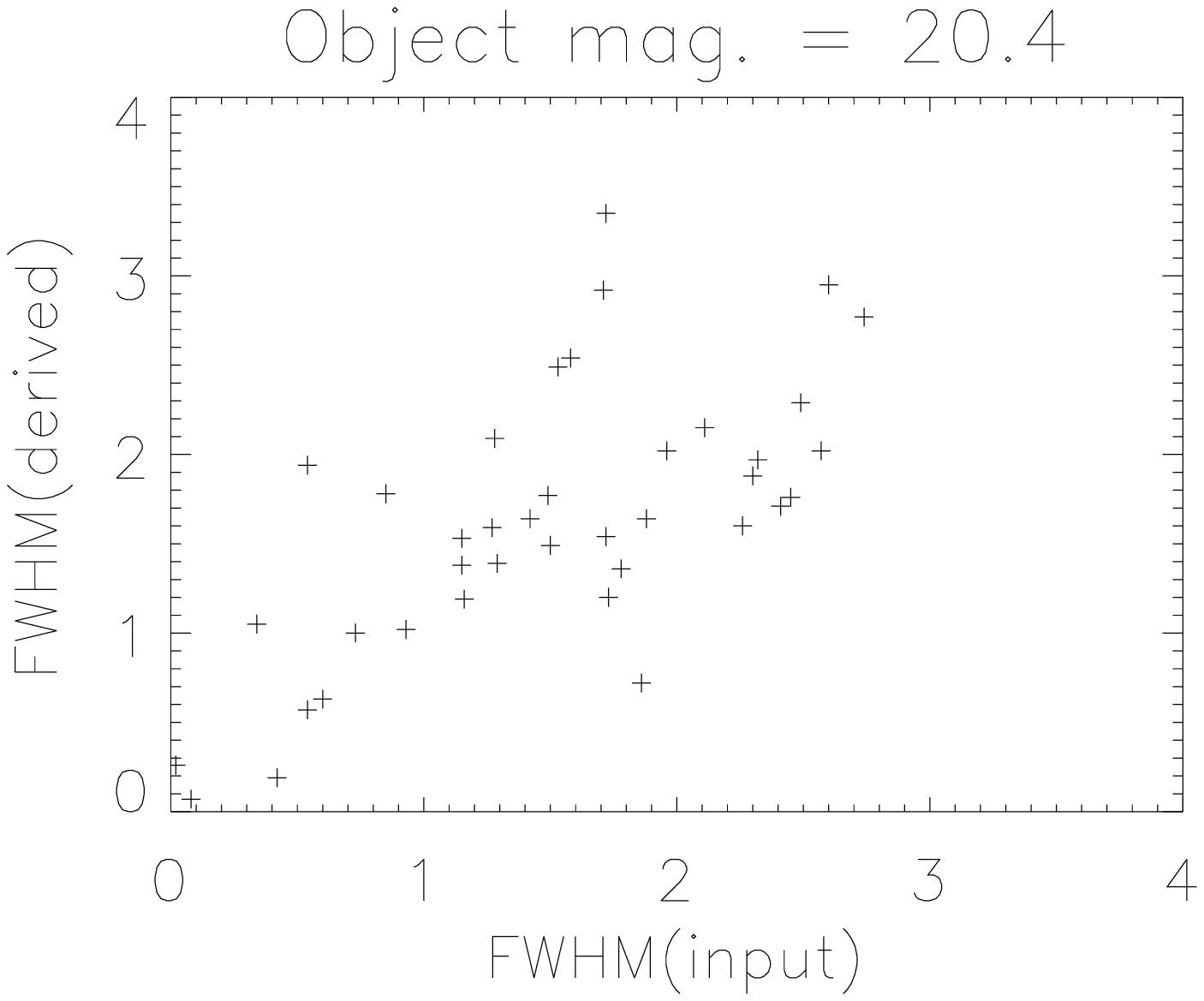}
\end{minipage}
\begin{minipage}{43mm}
\epsfxsize=43mm
\epsfbox[110 375 530 715]{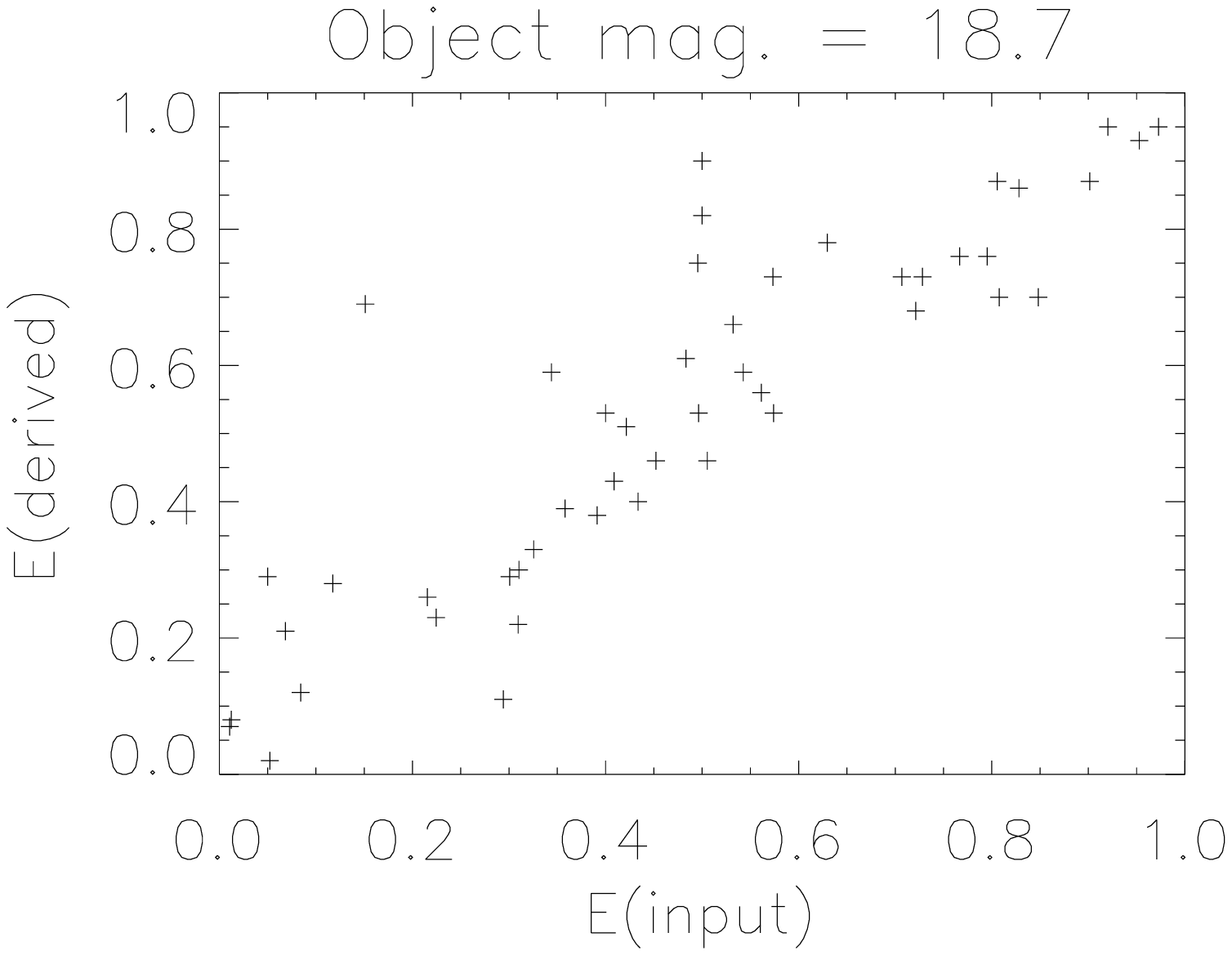}
\epsfxsize=43mm
\epsfbox[110 375 530 715]{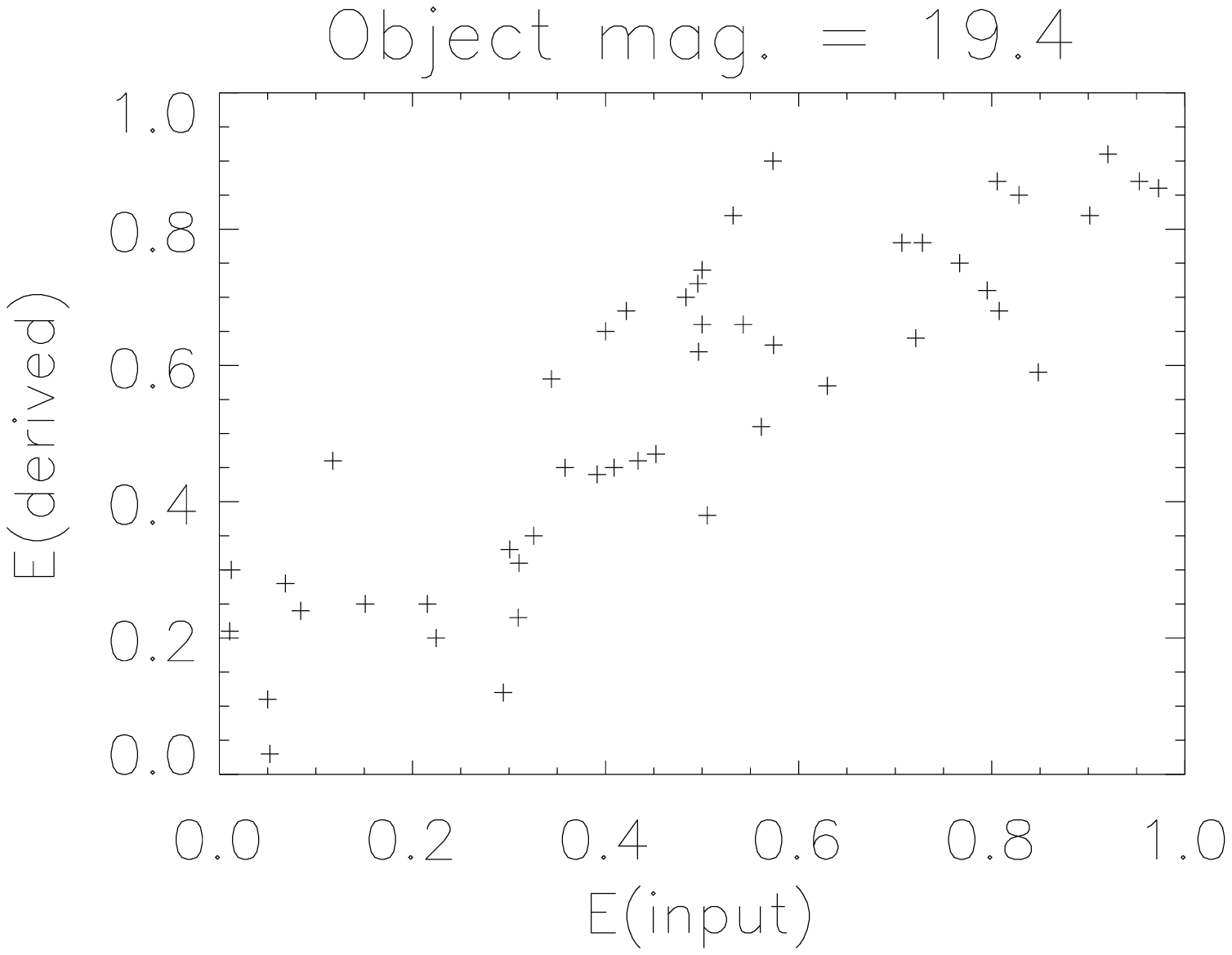}
\epsfxsize=43mm
\epsfbox[110 375 530 715]{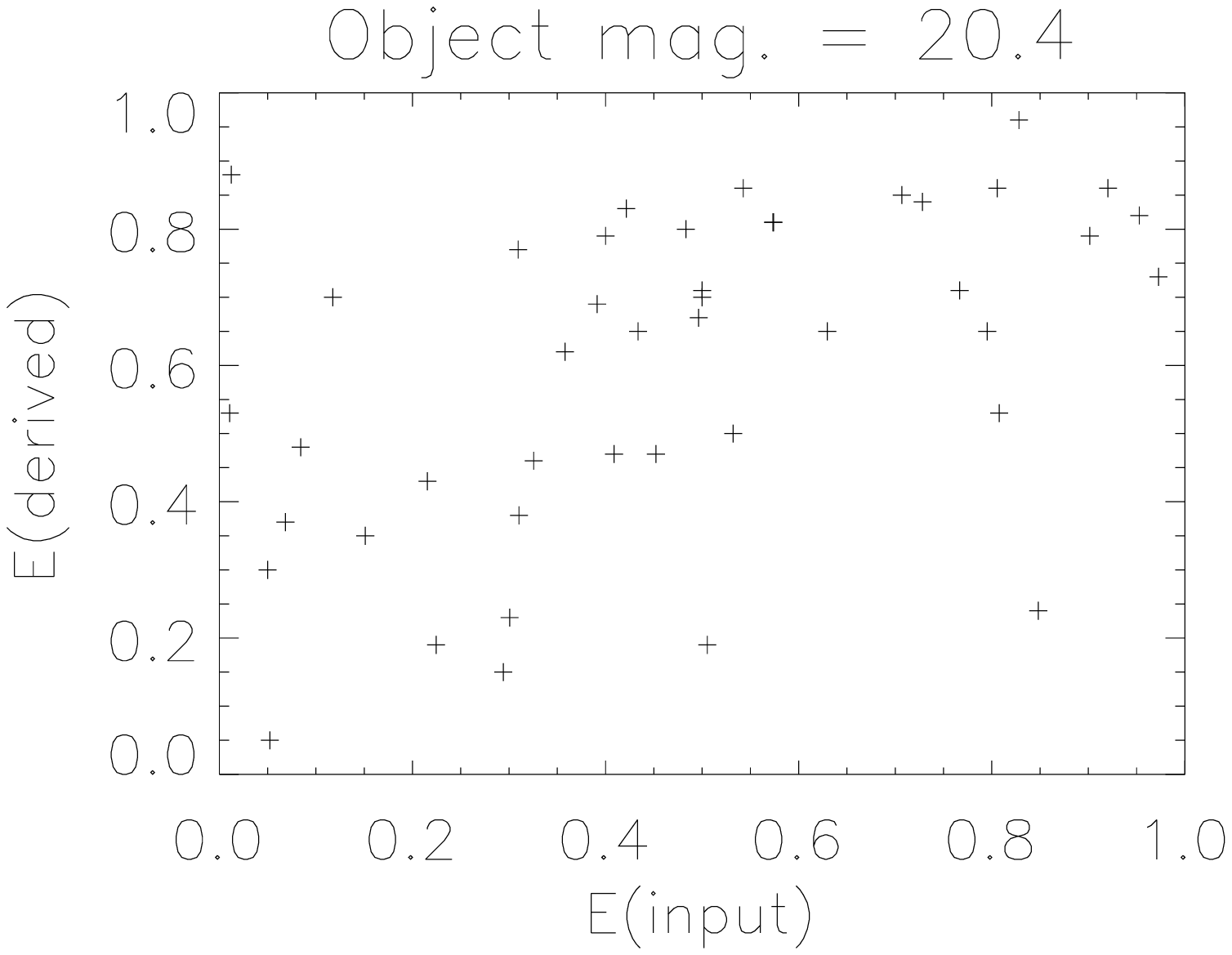}
\end{minipage}
\caption{
  Tests of {\bf ishape} using synthetic objects added to the \object{NGC~5236}
outer field. See Fig.~\ref{fig:ishape1} for details.
  \label{fig:ishape2}
}
\end{figure}

\begin{table}
\caption{
  Median S/N ratios for the data plotted in Figs.~\ref{fig:ishape1}
and \ref{fig:ishape2}.
\label{tab:sn}
}
\begin{tabular}{llll} \hline
Mag.        &  18.7  &  19.4  &  20.4 \\
Inner field &  142   &  79    &  36   \\
Outer field &  157   &  88    &  40   \\ \hline
\end{tabular}
\end{table}

  This was tested by generating a synthetic image with a number of 
objects with known shape parameters and then remeasuring them using 
{\bf ishape}. First, 49 test objects were generated by convolving the 
PSF measured on a V-band CCD image of the galaxy NGC~5236 with a number of 
MOFFAT15 models with major axis FWHMs in the range 0 - 3 pixels and axis 
ratios between 0 and 1. A synthetic image with all of the 49 test objects was 
then generated using {\bf mksynth}, and the synthetic image was finally added
to a section of the original image of NGC~5236. This procedure was repeated 
for synthetic objects of magnitudes 18.7, 19.4 and 20.4 at two positions
within NGC~5236 (see Fig.~\ref{fig:testfields}). Finally, {\bf ishape} was 
run on the test images, and the shape parameters derived by {\bf ishape} 
were compared with the input values.

  The results are shown in Figs.~\ref{fig:ishape1} (inner field) and 
\ref{fig:ishape2} (outer field).
The plots in the left column show the measured FWHM versus the input values,
and those in the right column show the measured axis ratio versus the
input values. The correlations are more tight for the outer
field, implying that the high noise level and stronger background
fluctuations in the bright disk near the centre of NGC~5236 limit 
{\bf ishape's} ability to reconstruct the
shape of the objects.   The median S/N ratios for the test objects 
calculated by {\bf ishape} are given in Table~\ref{tab:sn}, and as
expected the objects in the inner field have somewhat poorer S/N
ratios.  It is interesting to note that when sufficient signal is present 
the FWHM is recovered with quite high precision even down to very small 
values, below FWHM = 0.5 pixels. The FWHM of the PSF itself was 4 pixels, 
which means that objects with
sizes as small as about 10\% of that of the PSF can be recognised as
extended objects with good confidence. 

  From Table~\ref{tab:sn} and Figs.~\ref{fig:ishape1} and \ref{fig:ishape2} 
we estimate that the S/N ratio should be greater than about 50 in order to 
obtain reasonably accurate shape parameters, although there are probably also 
many other factors which affect the results, such as the presence of nearby
neighbours, crowding in general and the smoothness of the background.  
The axis ratio is generally somewhat more uncertain than the FWHM, although 
the scatter would decrease if the more compact objects were excluded from 
the plot. We will not discuss axis ratios further in this paper, but 
remark that the elongation might be used as a criterion to look for double
clusters.

\begin{figure}
\epsfxsize=85mm
\epsfbox[75 360 445 870]{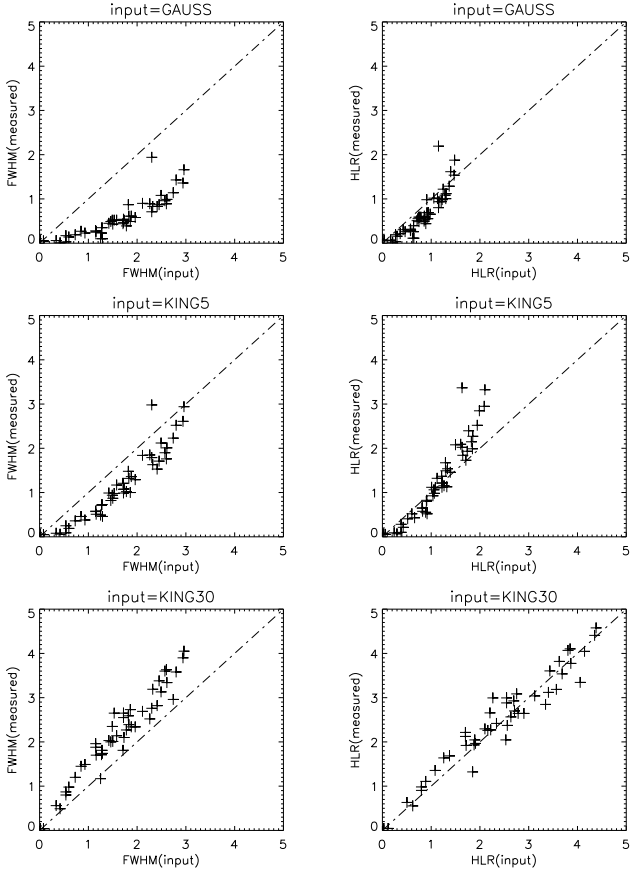}
\caption{\label{fig:itest}
  The FWHM and effective (half-light) radii derived from three test images 
by {\bf ishape}, using a MOFFAT15 profile. The test images were generated
using Gauss, King($c=5)$ and King($c=30$) profiles. While there are
obvious systematic errors in the measured FWHM values (left column) when 
applying a wrong model, the effective radii (right column) are in fact 
reproduced quite well. Sizes are in pixels.}
\end{figure}

  In the previous tests we had the advantage of knowing the intrinsic shape
of the synthetic clusters in advance. This will usually not be the case
in practice, so we should also examine how sensitive the derived cluster
sizes are to a particular choice of model. In this study we have used the
MOFFAT15 profile to model the clusters because of its similarity to
the models that were found to fit young LMC clusters by Elson et al. 
(\cite{elson1987}), but other choices might be as good. In particular, the 
classical models by King (\cite{king1962}) are known to fit galactic 
globular clusters very well. 
  We therefore generated another set of synthetic images in the same way as 
for the previously described {\bf ishape} tests, but now with Gaussian, 
King($c=5$) and King($c=30$) input models. The cluster sizes were then 
remeasured using {\bf ishape} and the MOFFAT15 model. The experiment was 
carried out only for the outer NGC~5236 field, and for one set of images 
(corresponding to the brightest set used in the previous tests).

  The results of this test are shown in Fig.~\ref{fig:itest} for FWHM
(left) and half-light radii (right). Though some systematic model dependencies
are evident for the FWHM values, half-light radii are reproduced
quite well by the MOFFAT15 model, regardless of the input model. A related
result was obtained by Kundu \& Whitmore (\cite{kw1998}) who found that
effective radii derived by fits to a King model were quite insensitive to the
adopted concentration parameter.  We may thus conclude that the derived 
effective radii are not very sensitive to the choice of model, and even if 
the true cluster profiles are closer to King profiles the choice of the 
MOFFAT15 model will not introduce any large systematic errors. 

\section{Completeness corrections}
\label{sec:compl}

\begin{figure}
\epsfxsize=88mm
\epsfbox[75 370 550 720]{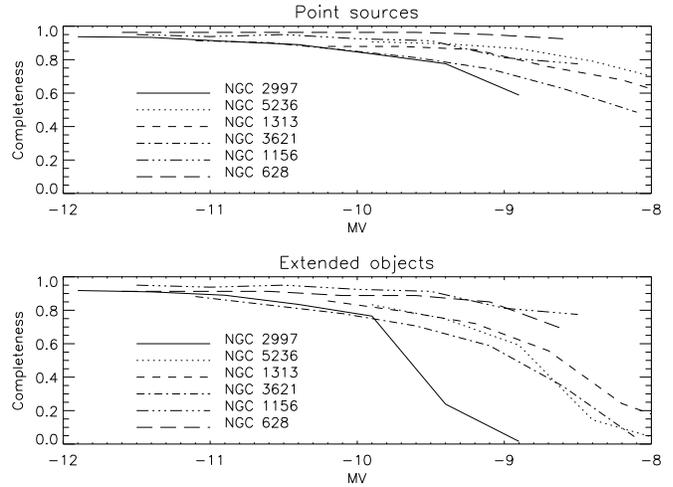}
\caption{
  Completeness corrections as a function of absolute visual magnitude
$M_V$ in six of the most
cluster-rich galaxies in the sample of galaxies in Paper1. The upper panel
shows the corrections applicable to point sources, while the lower
panel is for extended objects with $R_e = 20$ pc.
\label{fig:compl}
}
\end{figure}

  After the discussion of software tools we now return to YMCs. In Paper1 
the total numbers of clusters detected in each of the 21 galaxies in our 
sample were listed, and we also attempted to correct the raw numbers for 
completeness effects.  Here the completeness corrections will described in 
a bit more detail than in Paper1.

  The fraction of clusters that are not detected may be estimated by adding 
artificial clusters to the images using a programme like {\bf mksynth} and 
then checking how many of them are recovered by DAOFIND. This kind of 
completeness test is well documented in the literature for globular clusters 
in elliptical galaxies (Forbes \cite{forbes1996}, 
Kissler-Patig et al. \cite{kiss1996}) and also in 
other contexts. The artificial objects are usually added at 
random positions, which is a reasonable thing to do because this is also 
how the actual sources under study are distributed. 

  However, in our case the problem is somewhat different because YMCs are 
not distributed at random within their host galaxies. Instead, the clusters 
(in particular those in the ``blue'' group, i.e. the youngest ones) tend to 
be located in or near the spiral arms where they are much more likely to 
drown in background fluctuations, and hence a completeness test based on 
objects distributed at random is likely to underestimate the completeness 
correction. 

  Our approach has been to add the artificial objects ``close'' to the
detected objects, adding 5 clusters at random positions within a
radial range of 20 - 70 pixels from each detected cluster. The objects
were added to the $V$, $B$ and $U$ band frames using the {\bf mksynth} task,
and it was then required that they were refound by {\bf daofind} and that
{\bf phot} was able to obtain magnitudes in all three filters.
The completeness tests were carried out for magnitudes down to the 
$M_V = -8.5$ limit at intervals of 0.5 magnitudes, and a completeness 
correction was then calculated for each magnitude interval. The procedure 
was repeated twice, for point sources and for extended objects modeled
as a MOFFAT15 profile with an effective radius corresponding to 
$R_e = 20$ pc convolved with the PSF.  Finally, the number of
clusters actually detected in each magnitude interval was corrected by the
completeness correction calculated for that interval, and the corrected
numbers are given in Paper1 for the point-source as well for the
extended-source corrections. 

  Completeness tests were carried out only for the galaxies with more 
than 20 clusters.  Many of the cluster-poor galaxies contained only a 
handful or fewer clusters, and it makes little sense to attempt to apply 
completeness corrections to such small numbers.

  The estimated completenesses for sources in the 6 galaxies with more than 
20 clusters
are shown graphically in Fig.~\ref{fig:compl} as a function of absolute
visual magnitude. Note the striking difference between point sources 
(upper panel) and extended sources (lower panel), in particular in NGC~2997. 
The number of detections in NGC~2997 in the extended source experiment 
actually drops to 0 at $M_V = -9.0$, which indicates that in NGC~2997 we 
simply cannot make a realistic estimate of how many objects the galaxy 
contains down to a limit of $M_V = -8.5$. The other galaxies in our sample
were either more nearby, or they were observed at the NOT rather than the
1.54m telescope (and therefore with better image quality), so with the
exception of two more galaxies (NGC~7424 and NGC~1493, see Paper1) the
incompleteness corrections should be smaller than those for NGC~2997.

\section{Resolving YMCs in the galaxies}
\label{sec:resolv}

  The {\bf ishape} algorithm was applied to all the star clusters
identified as described in Sect.~\ref{subsec:identi}. For the shape model 
the MOFFAT15 profile was chosen because of its similarity with the models 
adopted by Elson et al. (\cite{elson1987}), and the PSF was generated 
using DAOPHOT. The cleaning radius was set to 2 pixels and the fitting 
radius to 4 pixels, and the CTRESH parameter was set to 2.

\begin{figure}
\begin{minipage}{43mm}
\epsfxsize=43mm
\epsfbox[95 365 540 720]{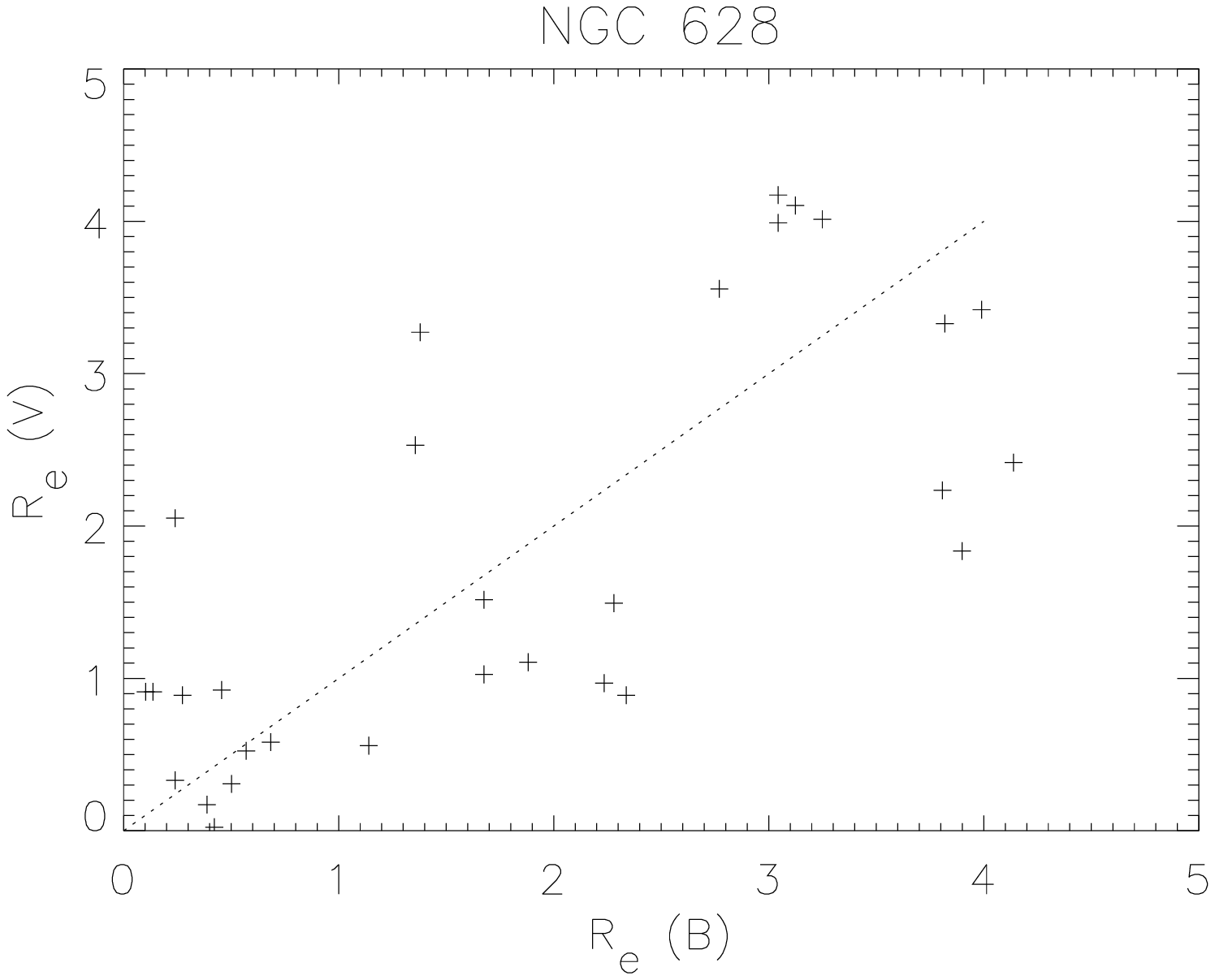}
\epsfxsize=43mm
\epsfbox[95 365 540 720]{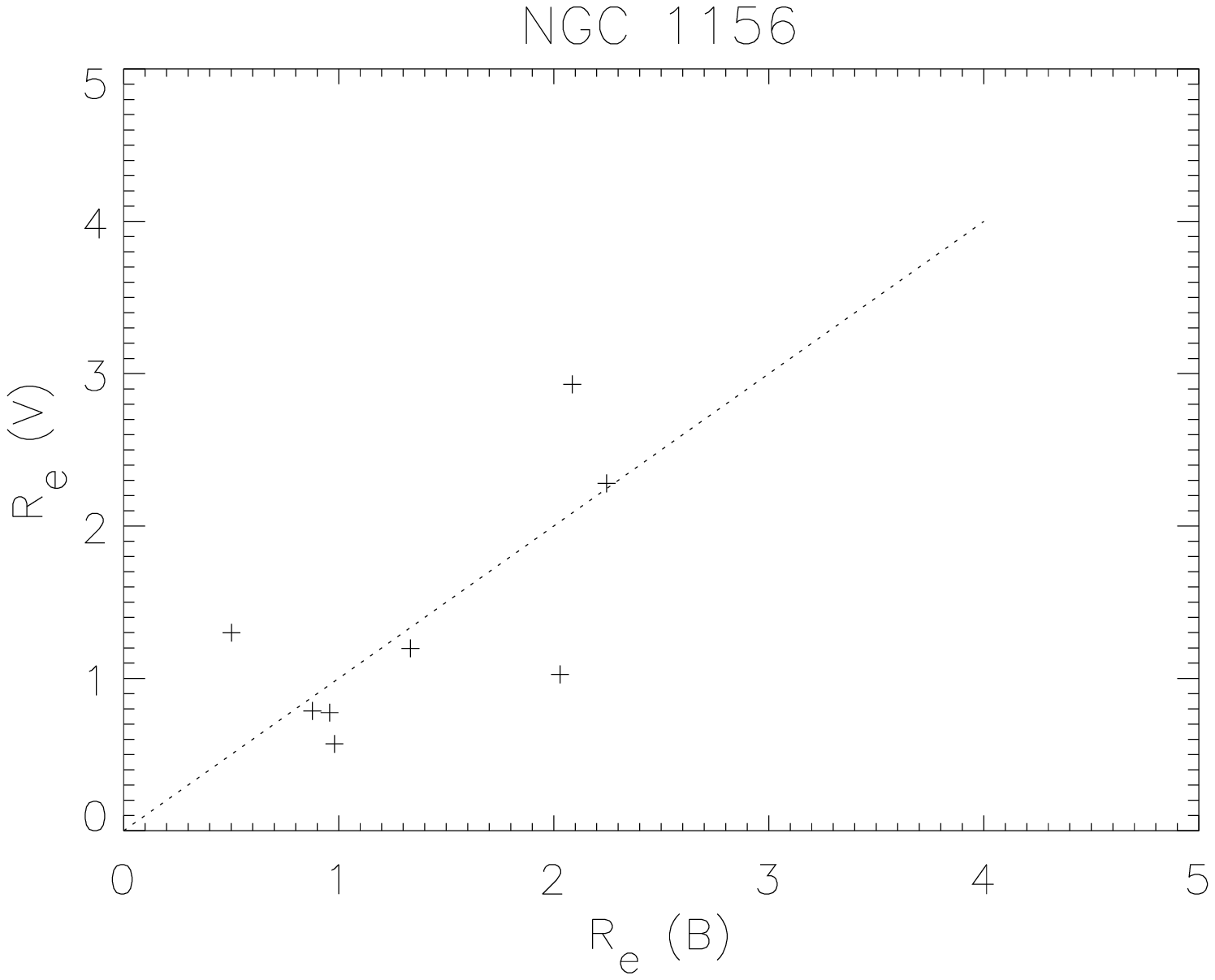}
\epsfxsize=43mm
\epsfbox[95 365 540 720]{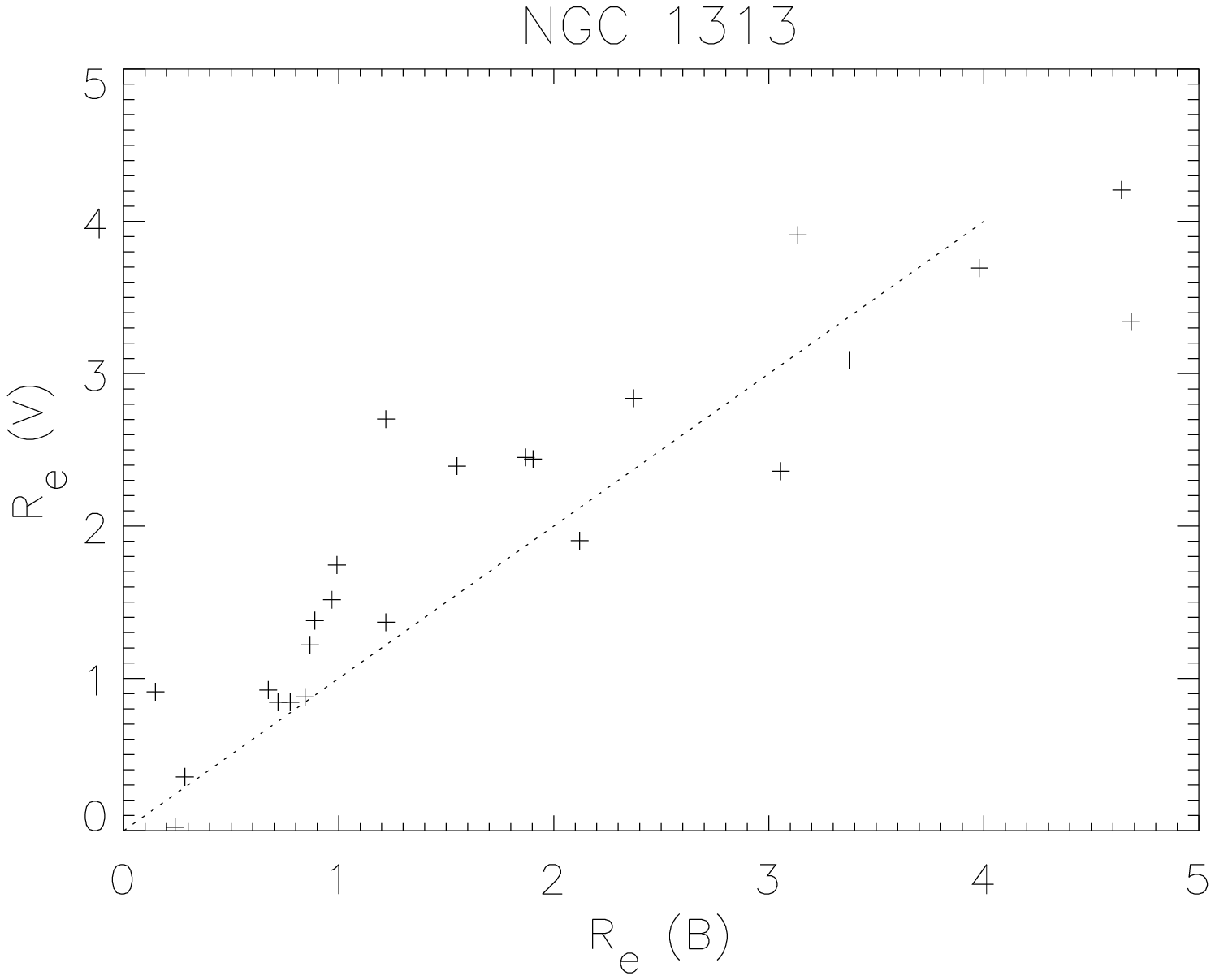}
\end{minipage}
\begin{minipage}{43mm}
\epsfxsize=43mm
\epsfbox[95 365 540 720]{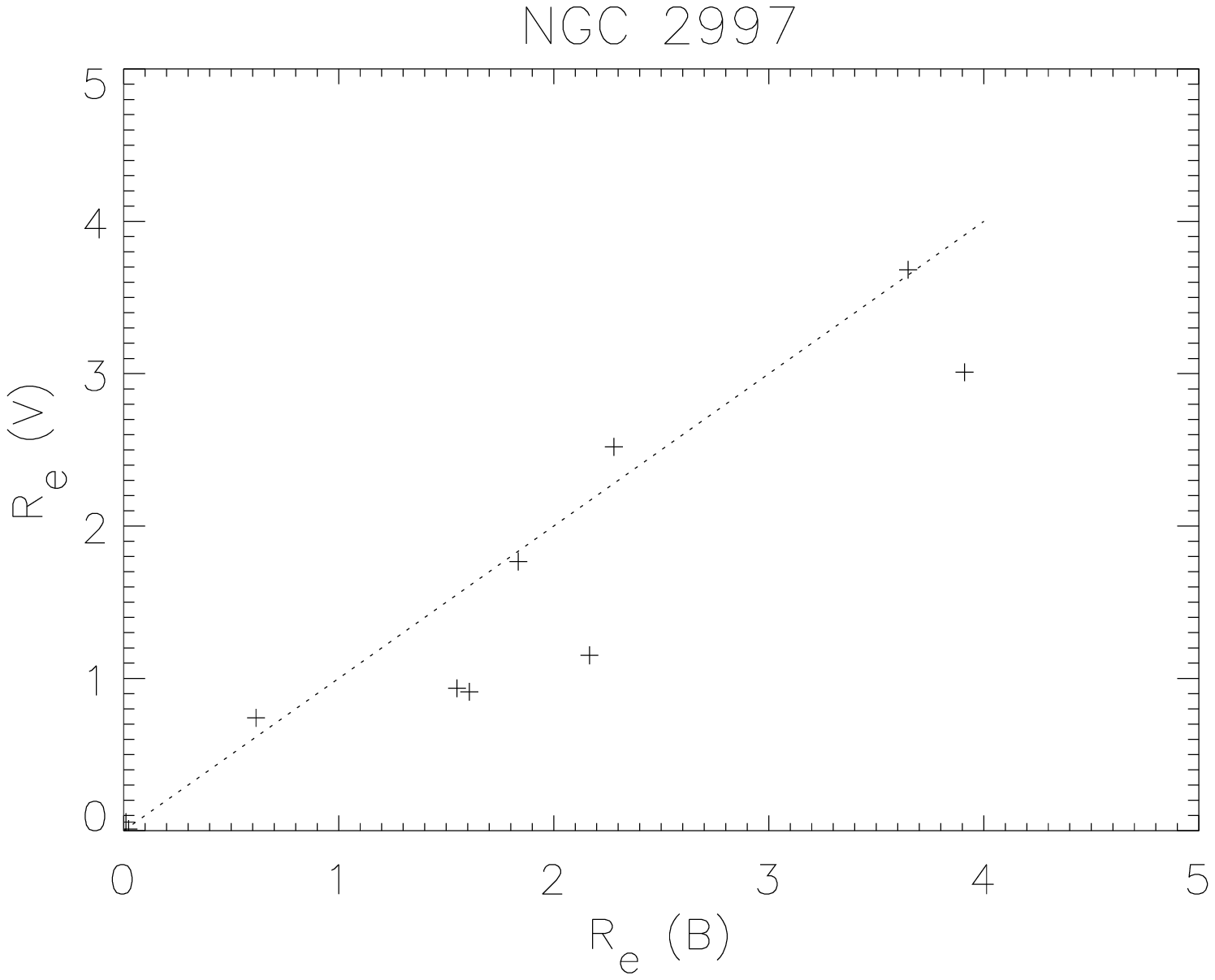}
\epsfxsize=43mm
\epsfbox[95 365 540 720]{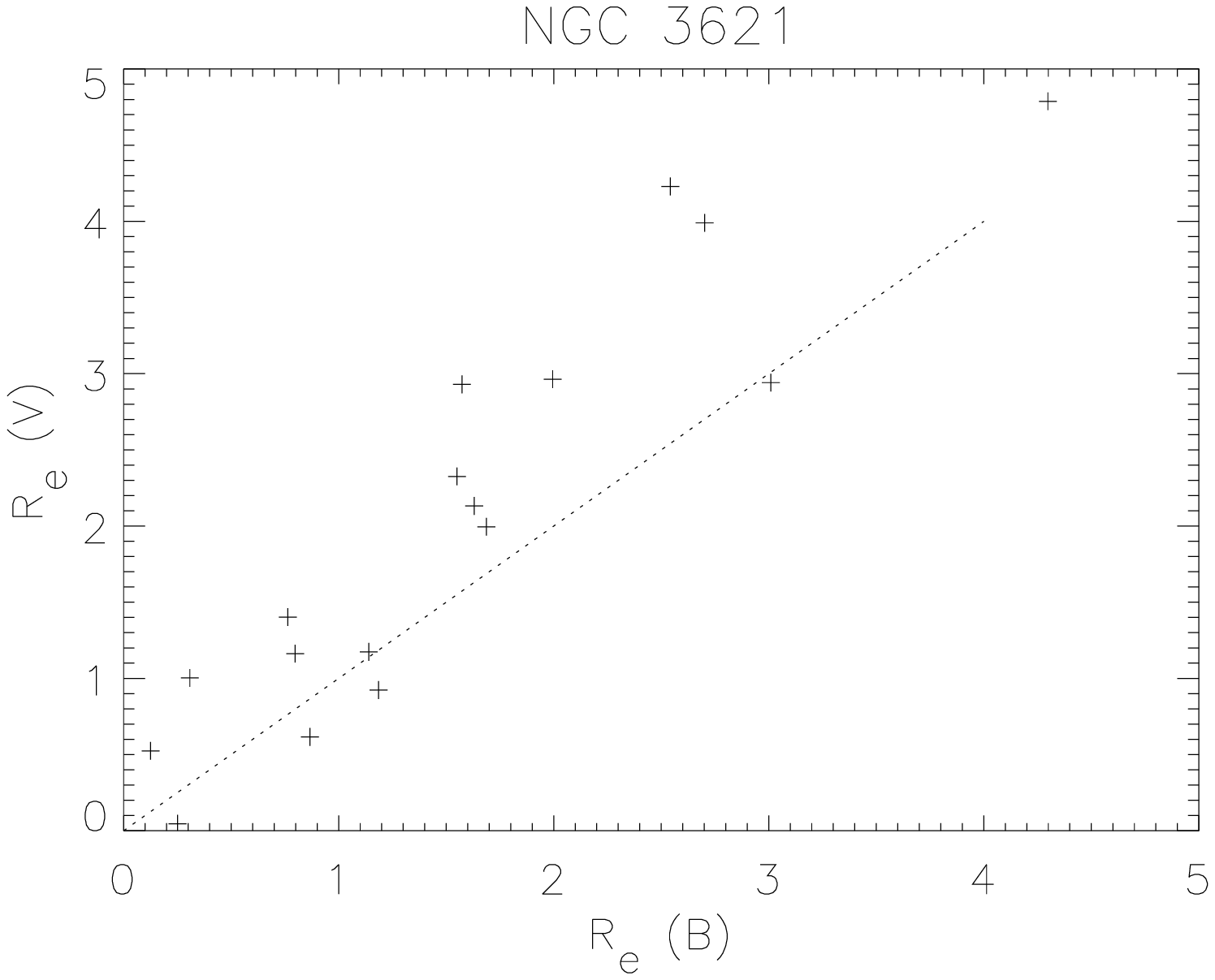}
\epsfxsize=43mm
\epsfbox[95 365 540 720]{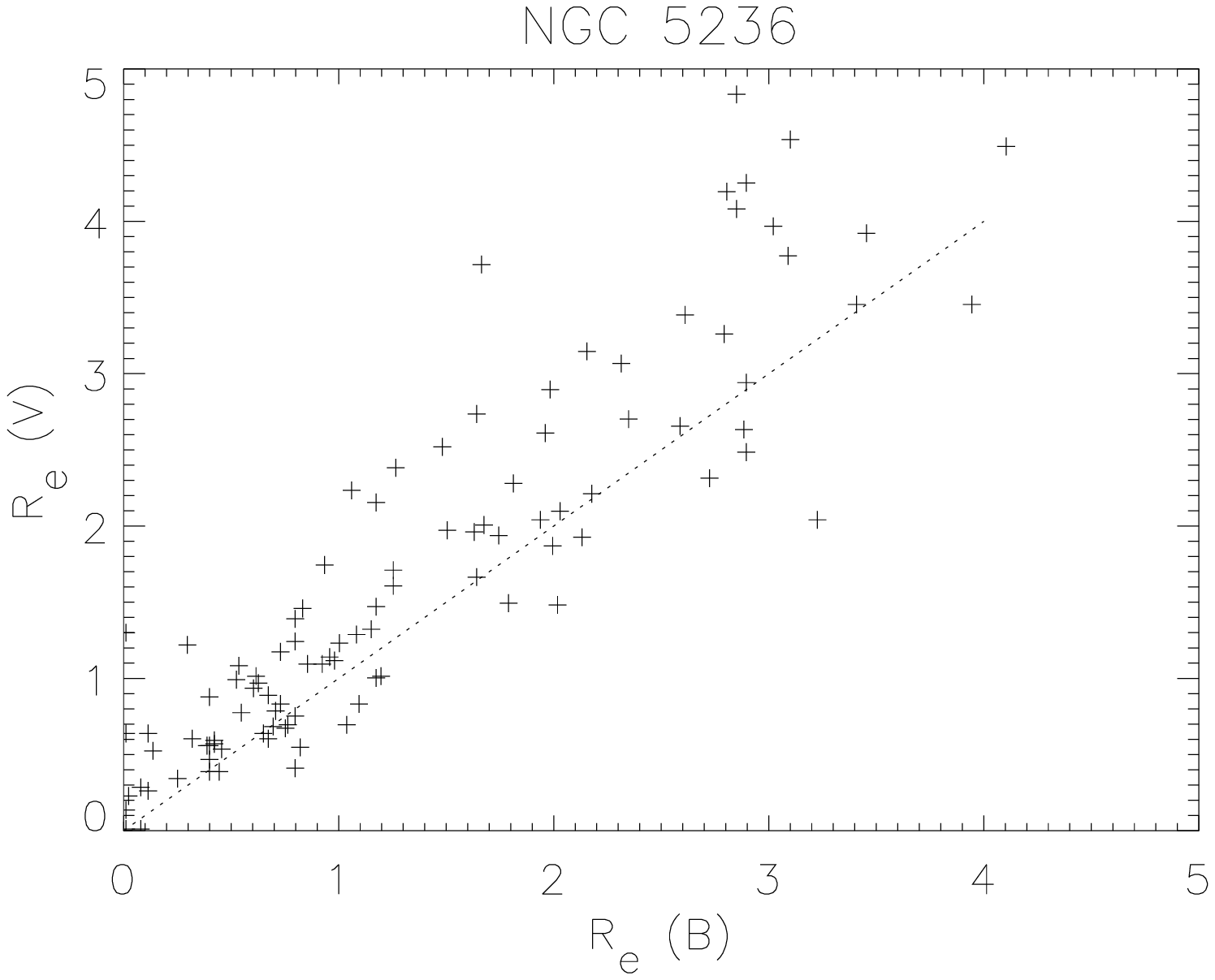}
\end{minipage}
\caption{
  The effective radii $R_e$ in pixel units of clusters measured on $V$ and 
$B$ band frames plotted against each other. Only clusters with S/N $>$ 50
are included in these plots.
  \label{fig:reff_bv}
}
\end{figure}

\begin{figure}
\epsfxsize=88mm
\epsfbox[95 365 545 535]{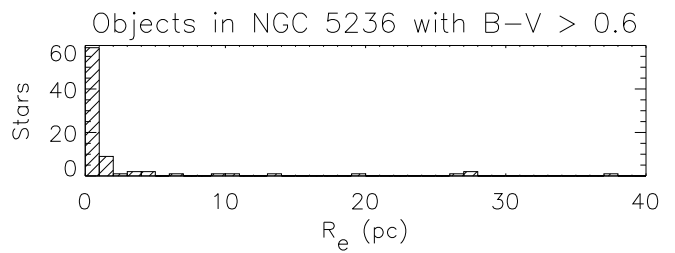}
\caption{
  Histogram of the effective radii of objects in NGC~5236 with
$V < 19.5$ and $B-V > 0.6$, assumed to be mostly foreground stars. Only 
objects in the region $300 < x < 1700$ pixels and $300 < y < 1700$ pixels 
were included. For easier comparison with Fig.~\ref{fig:szhist} the 
effective radii have been converted into parsec (1 pc $\approx$ 0.05\arcsec)
\label{fig:pthist}
}
\end{figure}

\begin{figure}
\epsfxsize=88mm
\epsfbox[85 365 545 720]{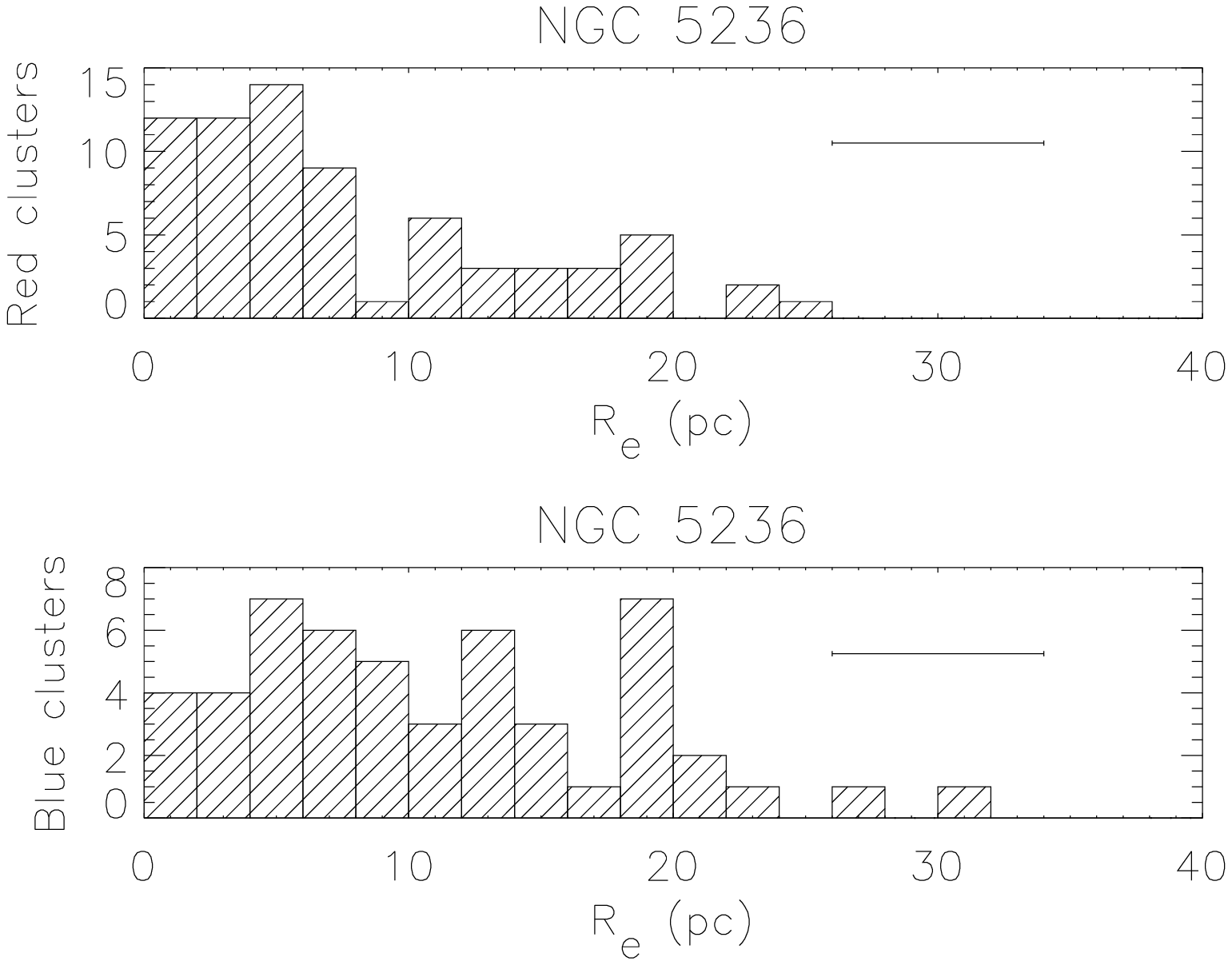}
\epsfxsize=88mm
\epsfbox[85 365 545 720]{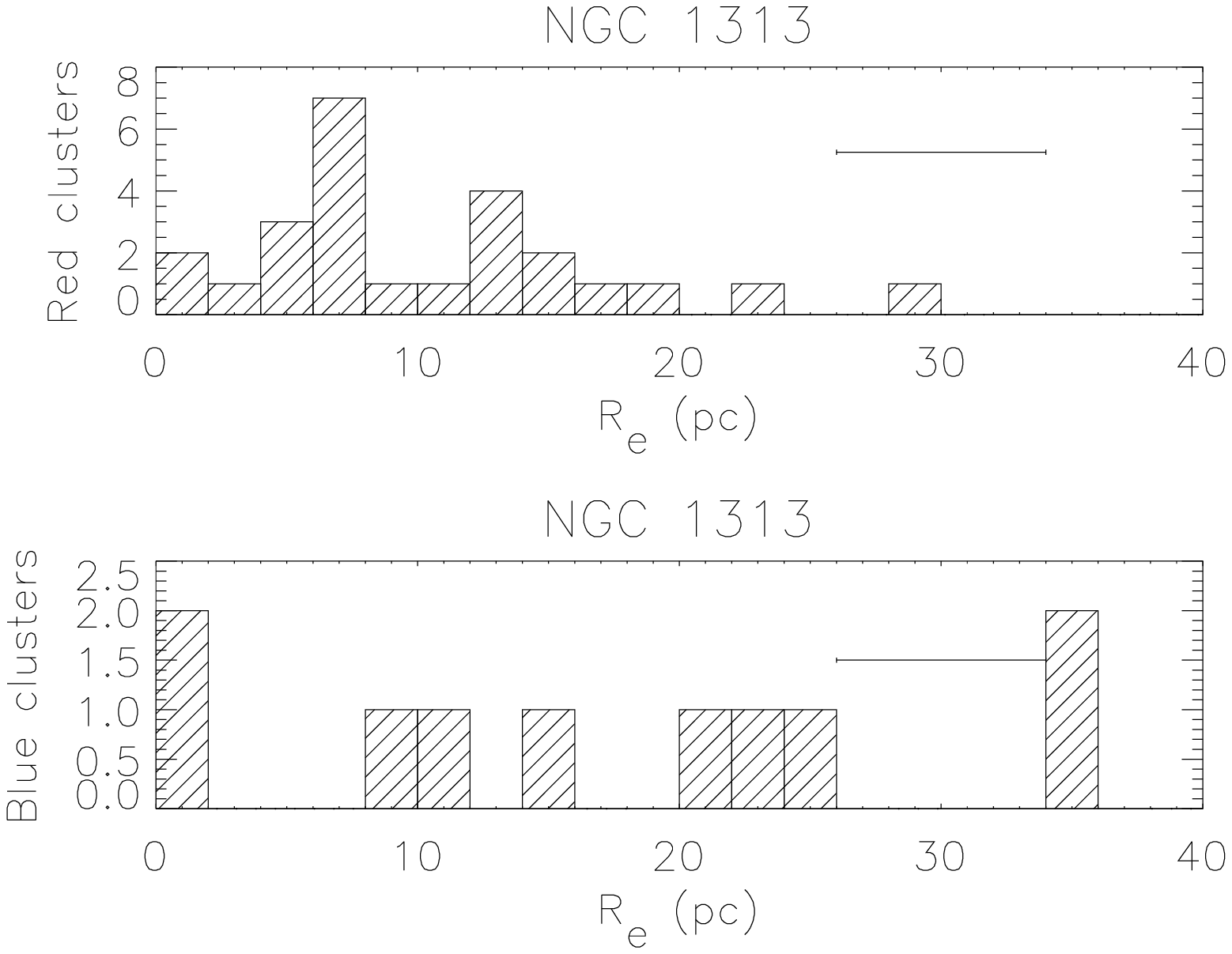}
\caption{
  Histograms of the effective radius $R_e$ for clusters in the ``red'' 
and ``blue'' groups in NGC~5236 (top) and NGC~1313 (bottom).  The 
effective radii in pixels were converted into parsec using the distances 
to the galaxies given in Paper1. The horizontal bar indicates the typical
uncertainty.
\label{fig:szhist}
}
\end{figure}

  The effective radii were derived both on $B$ and $V$ band images for the
6 most cluster-rich galaxies in order to check the results. In 
Fig.~\ref{fig:reff_bv} the effective
radii (in pixel units) derived by {\bf ishape} from $B$ and $V$ band frames 
are plotted against each other for objects with S/N $>$ 50, leaving out
about 20\% of the objects. If objects with 
lower S/N were included, the scatter was found to increase somewhat, although
the correlations in Fig.~\ref{fig:reff_bv} would remain evident. 

  As another test, radii were derived for sources in the NGC~5236 frame 
with \mbox{$B-V > 0.6$} and $V < 19.5$.  Most of these objects are 
expected to be foreground stars, and they are indeed distributed quite 
uniformly across the CCD frame.  Fig.~\ref{fig:pthist} shows the $R_e$ 
distribution for all such objects located within the central 
$1400 \times 1400$ pixels of the image, roughly corresponding to the area 
covered by the galaxy.  For convenience, the effective radii have been 
converted to parsec.  It can be seen from Fig.~\ref{fig:pthist} that 
{\bf ishape} finds $R_e < 1$ pc (0.05\arcsec) for the vast
majority of the objects with $B-V > 0.6$, in agreement with the assumption 
that they are in fact foreground stars. The fact that foreground stars appear
virtually unresolved by {\bf ishape} adds confidence to the algorithm's 
ability to recognise extended objects.

  The number of clusters in most of the galaxies are really too sparse that
it makes sense to discuss the $R_e$ distributions for individual galaxies. 
However, Fig.~\ref{fig:szhist} shows $R_e$ histograms for two of the most 
cluster-rich galaxies, NGC~5236 and NGC~1313.  These are also the two most 
nearby galaxies, at $m-M=27.84$ and $28.2$ respectively (see Paper1).  
Separate histograms are given for the ``red'' (U-B $> -0.4$) 
and ``blue'' (U-B $< -0.4$) cluster samples.  Several systematic effects 
might influence the observed $R_e$ distributions, for example the fact that 
the most extended clusters are detected with a much lower efficiency, and 
one should be careful not to overinterpret the data in Fig.~\ref{fig:szhist}.
Nevertheless, the objects in the ``blue'' sample generally seem to be
less concentrated, which might imply that many of the blue objects are 
loosely bound associations that do not survive long, rather than bound 
clusters. It is emphasised, however, that this statement could only be 
justified by a more detailed study of the structure of the objects, something 
that is not feasible from ground-based observations. 
  Fig.~\ref{fig:szhist} also seems to indicate that many of the objects 
in NGC~5236 are more concentrated than those in NGC~1313, although the 
statistics are poor for the latter galaxy. 

  From the discussion in Sect.~\ref{subsec:itest} and Fig.~\ref{fig:reff_bv}
the uncertainty on the cluster radii is estimated to be about 0.5 pixels.
1 DFOSC pixel ($0.4\arcsec$) corresponds to about 7 pc at the distance of 
NGC~5236 and NGC~1313, so we expect an uncertainty of about 4 pc on the 
individual cluster radii in Fig.~\ref{fig:szhist}, as indicated by
the horizontal error bars. The fact that we do not 
observe any lower limit in the $R_e$ distribution may be an effect of the 
limited resolution, causing clusters with larger radii to scatter into the 
low $R_e$ region, but the lower $R_e$ limit in our sample cannot be higher 
than a few pc.

  How do the sizes of the clusters discussed here compare with other 
results?  Accurate measurements of cluster sizes are available only in the
Milky Way and in the LMC.  Most globular clusters in the Milky Way have 
$R_e < 10$ pc with a peak in the $R_e$ distribution at about 3 pc, although 
clusters with $R_e$ up to $\sim 30$ pc exist (Harris \cite{harris1996}). 
Elson et al. (\cite{elson1987}) found typical half-mass radii for young 
LMC clusters in the range 5 - 15 pc, corresponding to $R_e$ values between 
4 and 12 pc (Spitzer \cite{spitz1987}).

  Effective radii have been estimated also for young clusters in the Antennae 
by Whitmore \& Schweizer (\cite{ws1995}, WS95) using HST/WFPC data, and in 
the merger remnant NGC~7252 (Whitmore et al. \cite{whit93}), in both cases by 
assuming gaussian cluster profiles. WS95 found an average $R_e$ of 12 pc for 
the Antennae clusters, but new WFPC/2 data lead to a revised 
mean value of $R_e = 4\pm 1$ pc (Whitmore et al. \cite{whit99}). In the case of 
NGC~7252, Whitmore et al.  (\cite{whit93}) 
found a mean effective radius of 9.9 pc for clusters located at distances 
larger than 3.5\arcsec\ from the centre of NGC~7252. {\"O}stlin et al. 
(\cite{ostlin1998}) used the model profile of Lugger et al. (\cite{lug1995}) 
to estimate {\it core} radii 
(one half times the FWHM) for YMCs in the blue compact galaxy ESO 338-IG04,
and found a distribution that peaked at 2.5 pc with all clusters having core
radii less than 10 pc. The Lugger et al. (\cite{lug1995}) profile does not 
have a well-defined effective radius, but {\"O}stlin et al. (\cite{ostlin1998}) 
also found that if the clusters were instead fitted with gaussian profiles 
(for which the core radius equals $R_e$) the core radii were about 2 times 
larger than for the Lugger et al. (\cite{lug1995}) profiles, so their $R_e$ 
distribution should peak at about 5 pc and have an upper limit at 20 pc. 

  Our observed $R_e$ distributions thus seem to resemble those observed for
YMCs in other galaxies quite well. We note, however, that the effective radii 
have been estimated by a variety of methods in the different studies, and 
this could account for some of the differences.  The effective radii for the 
YMCs in our sample are also comparable with those of Galactic globular 
clusters and young LMC clusters.

  The velocity dispersion in a star cluster or OB association is typically 
of the order of 1 - 10 km/sec, so an unbound object will expand by 1 - 10 
pc / Myr. A U-B colour of $-0.4$ corresponds to an age of about 50 Myr 
(Girardi et al. \cite{girardi1995}), so most of the objects in the 
``red'' group would have expanded to sizes much larger than what is observed 
and would have
effectively disappeared if they were not gravitationally bound.  Like all 
star clusters they will eventually be subject to dynamical erosion, 
but this acts on much longer timescales and the lifetimes are difficult to
estimate without a detailed knowledge of the morphology of individual
clusters and their orbits within the host galaxies.

\section{Summary and conclusions}

  The data reductions applied to the sample of 21 galaxies presented in 
Paper1 have been extensively discussed, in particular the algorithm 
{\bf ishape} which is a useful tool for the analysis of compact objects. 
Tests show that {\bf ishape} is able to recognise extended objects with 
good confidence, down to a FWHM of about 1/10 of the FWHM of the PSF itself.

  Completeness tests of the cluster samples of the 6 most cluster-rich
galaxies were carried out using {\bf mksynth}, and it was demonstrated 
that the completeness depends strongly on the extent of the objects. 

  Using {\bf ishape}, effective radii of YMCs were obtained, and the 
$R_e$ distributions in NGC~1313 and NGC~5236 were studied in more detail.
We found effective radii of the YMCs in these two galaxies in the range
0 - 20 pc, in good agreement with results obtained for YMCs in other 
galaxies, and also similar to those of old globular clusters in the Milky 
Way and YMCs in the LMC.  Some of the objects with \mbox{U-B}$<-0.4$ may be 
loosely bound associations, while objects with \mbox{U-B}$>-0.4$ (older than 
$\sim 50$ Myr) are orders of magnitude older than their crossing times,
and we may thus conclude that they are genuine clusters. Many of them are 
significantly more massive than any open cluster in the Milky Way, and may 
well evolve into objects resembling the old globular clusters seen in the 
halo of the Milky Way and other galaxies today.

\begin{acknowledgements}
  This research was supported by the Danish Natural Science Research Council
through its Centre for Ground-Based Observational Astronomy. The valuable
comments of the referee, Dr. Jon Holtzman, which helped improve this
paper are appreciated.  This research has made use of the NASA/IPAC 
Extragalactic Database (NED) which is operated by the Jet Propulsion 
Laboratory, California Institute of Technology, under contract with the 
National Aeronautics and Space Administration. 
\end{acknowledgements}


\begin{table*}
\caption{
  Data for all the clusters identified in this study. Column 1 identifies
each cluster by the name of its host galaxy and a number. E.g. ``n1313-221''
refers to a cluster in the galaxy NGC~1313. The column $(x,y)$ gives the 
pixel coordinates of each cluster, $M_V$ is the absolute visual magnitude 
and $U-B$, $B-V$ and $V-I$ are the broad-band colours. All photometric data 
are corrected for galactic foreground extinction. $R_e$ is the effective 
radius in pc, as estimated by {\bf ishape} using a MOFFAT15 model.
  \label{tab:clusters}
}
\end{table*}

\setcounter{table}{3}
\begin{table*}
\caption{(continued)}
\begin{tabular}{lrrrrrrrr} \\ \hline
  ID      &  $(x,y)$  & $\alpha$(2000.0) & $\delta$(2000.0) & $M_V$ & U-B & B-V  & V-I  & $R_e$ \\ \hline
n45-121 & 645,685 & 0:14:15.10 & -23:13:24.9 & -9.94 & -0.64 & 0.34 & 0.68 & 32.8 \\
n45-271 & 1104,1034 & 0:14:01.74 & -23:11:12.8 & -8.77 & -0.14 & 0.43 & 0.66 & 4.7 \\
n45-341 & 1030,1203 & 0:14:03.72 & -23:10:05.8 & -8.51 & -0.31 & 0.09 & 0.33 & 27.6 \\
n247-508 & 949,1057 & 0:47:09.29 & -20:45:30.2 & -8.63 & -0.28 & 0.33 & 0.58 & 37.7 \\
n247-871 & 1327,1715 & 0:46:58.25 & -20:41:16.0 & -10.24 & -0.67 & 0.02 & 0.39 & 4.9 \\
n247-899 & 985,1784 & 0:47:07.76 & -20:40:45.5 & -9.38 & -0.06 & 0.16 & 0.46 & 6.5 \\
\hline
\end{tabular}
\end{table*}

\setcounter{table}{3}
\begin{table*}
\caption{(continued)}
\begin{tabular}{lrrrrrrrr} \\ \hline
  ID      &  $(x,y)$  & $\alpha$(2000.0) & $\delta$(2000.0) & $M_V$ & U-B & B-V  & V-I  & $R_e$ \\ \hline
n1156-442 & 1372,1042 & 2:59:42.44 & 25:14:08.6 & -9.52 & -0.49 & -0.01 & 0.29 & 15.2 \\
n1156-476 & 1329,1067 & 2:59:43.03 & 25:14:13.2 & -9.53 & -0.41 & 0.09 & 0.40 & 3.7 \\
n1156-496 & 1426,1085 & 2:59:41.70 & 25:14:16.8 & -9.68 & -0.65 & 0.10 & 0.47 & 6.4 \\
n1156-513 & 1404,1099 & 2:59:42.01 & 25:14:19.3 & -10.01 & -0.53 & 0.12 & 0.40 & 7.1 \\
n1156-549 & 1455,1125 & 2:59:41.31 & 25:14:24.3 & -9.74 & -0.12 & 0.25 & 0.48 & 20.7 \\
n1156-554 & 1269,1129 & 2:59:43.87 & 25:14:24.6 & -9.74 & -0.42 & 0.33 & 0.64 & 24.0 \\
\hline
\end{tabular}
\end{table*}

\setcounter{table}{3}
\begin{table*}
\caption{(continued)}
\begin{tabular}{lrrrrrrrr} \\ \hline
  ID      &  $(x,y)$  & $\alpha$(2000.0) & $\delta$(2000.0) & $M_V$ & U-B & B-V  & V-I  & $R_e$ \\ \hline
n1313-783 & 807,1388 & 3:18:26.75 & -66:27:24.5 & -9.70 & -0.08 & 0.28 & 0.59 & 6.1 \\
n1313-860 & 323,1857 & 3:18:57.76 & -66:24:16.6 & -9.68 & -0.66 & 0.29 & 0.73 & 1.3 \\
n2403-250 & 944,271 & 7:36:54.14 & 65:33:29.9 & -8.76 & -0.30 & 0.25 & 0.54 & 10.0 \\
n2403-850 & 1108,529 & 7:36:49.31 & 65:34:18.3 & -9.25 & -0.31 & 0.27 & 0.37 & 0.9 \\
n2403-939 & 708,568 & 7:37:01.33 & 65:34:24.6 & -8.80 & -0.27 & 0.33 & 0.38 & 1.1 \\
n2403-1157 & 575,635 & 7:37:05.34 & 65:34:36.8 & -8.72 & 0.07 & 0.44 & 0.42 & 1.6 \\
\hline
\end{tabular}
\end{table*}

\setcounter{table}{3}
\begin{table*}
\caption{(continued)}
\begin{tabular}{lrrrrrrrr} \\ \hline
  ID      &  $(x,y)$  & $\alpha$(2000.0) & $\delta$(2000.0) & $M_V$ & U-B & B-V  & V-I  & $R_e$ \\ \hline
n2997-838 & 1260,1350 & 9:45:33.33 & -31:09:59.6 & -10.27 & -0.41 & 0.02 & 0.36 & 8.2 \\
n2997-999 & 847,1660 & 9:45:45.77 & -31:07:54.6 & -9.45 & -0.16 & 0.33 & 0.31 & 2.3 \\
n2997-1013 & 1583,1682 & 9:45:23.27 & -31:07:51.7 & -10.71 & -0.25 & 0.16 & 0.32 & 34.0 \\
n3184-105 & 1217,464 & 10:18:06.71 & 41:25:58.8 & -8.81 & -0.20 & 0.11 & 0.44 & 6.1 \\
n3184-353 & 824,771 & 10:18:11.71 & 41:24:44.7 & -10.49 & -0.56 & -0.01 & 0.54 & 15.4 \\
n3184-383 & 1635,823 & 10:18:12.71 & 41:27:16.4 & -9.68 & -0.52 & 0.17 & 0.52 & 22.6 \\
\hline
\end{tabular}
\end{table*}

\setcounter{table}{3}
\begin{table*}
\caption{(continued)}
\begin{tabular}{lrrrrrrrr} \\ \hline
  ID      &  $(x,y)$  & $\alpha$(2000.0) & $\delta$(2000.0) & $M_V$ & U-B & B-V  & V-I  & $R_e$ \\ \hline
n3621-1515 & 1022,1311 & 11:18:15.35 & -32:46:55.2 & -10.27 & -0.32 & 0.18 & 0.49 & 28.8 \\
n3621-1526 & 1092,1329 & 11:18:13.15 & -32:46:48.7 & -10.56 & -0.78 & 0.11 & 0.50 & 9.8 \\
n3621-1527 & 1004,1330 & 11:18:15.90 & -32:46:47.6 & -10.20 & -0.43 & 0.21 & 0.36 & 1.6 \\
n3621-1534 & 1012,1336 & 11:18:15.64 & -32:46:45.3 & -10.42 & -0.46 & 0.10 & 0.31 & 12.1 \\
n3621-1776 & 1462,1732 & 11:18:01.37 & -32:44:13.5 & -8.65 & -0.25 & 0.29 & 0.24 & 6.3 \\
n4395-469 & 365,1480 & 12:25:54.59 & 33:30:47.6 & -9.06 & -0.08 & 0.27 & 0.39 & 2.1 \\
\hline
\end{tabular}
\end{table*}

\setcounter{table}{3}
\begin{table*}
\caption{(continued)}
\begin{tabular}{lrrrrrrrr} \\ \hline
  ID      &  $(x,y)$  & $\alpha$(2000.0) & $\delta$(2000.0) & $M_V$ & U-B & B-V  & V-I  & $R_e$ \\ \hline
n5236-531 & 1395,857 & 13:36:49.08 & -29:53:16.3 & -9.61 & -0.38 & 0.08 & 0.41 & 6.1 \\
n5236-548 & 493,867 & 13:37:16.29 & -29:53:05.2 & -9.61 & -0.11 & 0.21 & 0.33 & 5.5 \\
n5236-549 & 795,868 & 13:37:07.18 & -29:53:07.3 & -10.14 & -0.31 & 0.23 & 0.41 & 16.1 \\
n5236-552 & 1052,869 & 13:36:59.42 & -29:53:08.9 & -9.42 & -0.33 & 0.39 & 0.80 & 25.3 \\
n5236-555 & 1428,871 & 13:36:48.07 & -29:53:11.1 & -9.47 & -0.32 & 0.21 & 0.55 & 5.4 \\
n5236-572 & 1435,883 & 13:36:47.85 & -29:53:06.4 & -8.65 & -0.10 & 0.44 & 0.90 & 0.6 \\
\hline
\end{tabular}
\end{table*}

\setcounter{table}{3}
\begin{table*}
\caption{(continued)}
\begin{tabular}{lrrrrrrrr} \\ \hline
  ID      &  $(x,y)$  & $\alpha$(2000.0) & $\delta$(2000.0) & $M_V$ & U-B & B-V  & V-I  & $R_e$ \\ \hline
n5236-1201 & 1547,1230 & 13:36:44.27 & -29:50:51.1 & -8.65 & -0.33 & 0.26 & 0.67 & 5.4 \\
n5236-1227 & 1692,1249 & 13:36:39.88 & -29:50:44.7 & -8.55 & -0.12 & 0.18 & 0.45 & 3.9 \\
n5236-1234 & 873,1255 & 13:37:04.58 & -29:50:35.9 & -10.17 & -0.63 & -0.04 & 0.41 & 7.2 \\
n5236-1235 & 758,1255 & 13:37:08.05 & -29:50:35.0 & -10.98 & -0.74 & -0.03 & 0.29 & 13.2 \\
n5236-1237 & 1396,1256 & 13:36:48.81 & -29:50:39.7 & -8.52 & -0.20 & 0.28 & 0.51 & 0.1 \\
n5236-1242 & 1375,1258 & 13:36:49.44 & -29:50:38.7 & -8.66 & -0.33 & 0.38 & 0.51 & 12.2 \\
\hline
\end{tabular}
\end{table*}

\setcounter{table}{3}
\begin{table*}
\caption{(continued)}
\begin{tabular}{lrrrrrrrr} \\ \hline
  ID      &  $(x,y)$  & $\alpha$(2000.0) & $\delta$(2000.0) & $M_V$ & U-B & B-V  & V-I  & $R_e$ \\ \hline
n5585-271 & 1064,1080 & 14:19:47.91 & 56:43:54.5 & -9.66 & -0.38 & 0.13 & 0.39 & 2.6 \\
n5585-284 & 767,1090 & 14:19:48.09 & 56:42:59.2 & -8.59 & -0.09 & 0.29 & 0.68 & 4.0 \\
n5585-285 & 1002,1090 & 14:19:48.12 & 56:43:43.0 & -10.82 & -0.55 & -0.05 & 0.32 & 19.2 \\
n5585-410 & 1563,1382 & 14:19:54.82 & 56:45:27.1 & -9.33 & -0.35 & 0.08 & 0.28 & 4.0 \\
n6744-359 & 618,482 & 19:10:09.63 & -63:54:45.1 & -9.36 & -0.28 & 0.15 & 0.41 & 2.0 \\
n6744-480 & 1005,613 & 19:09:46.43 & -63:53:57.0 & -9.59 & -0.64 & 0.13 & 0.42 & 40.4 \\
\hline
\end{tabular}
\end{table*}

\setcounter{table}{3}
\begin{table*}
\caption{(continued)}
\begin{tabular}{lrrrrrrrr} \\ \hline
  ID      &  $(x,y)$  & $\alpha$(2000.0) & $\delta$(2000.0) & $M_V$ & U-B & B-V  & V-I  & $R_e$ \\ \hline
n6946-1989 & 682,930 & 20:34:58.90 & 60:08:47.9 & -9.70 & -0.67 & -0.07 & -0.07 & 2.7 \\
n6946-2044 & 682,950 & 20:34:58.90 & 60:08:51.6 & -10.32 & -0.64 & 0.03 & 0.46 & 4.4 \\
n6946-2064 & 1423,956 & 20:34:40.40 & 60:08:54.2 & -9.98 & -0.92 & -0.13 & -0.09 & 6.5 \\
n6946-2074 & 1730,960 & 20:34:32.73 & 60:08:55.6 & -8.85 & -0.37 & 0.06 & 0.26 & 6.9 \\
n6946-2075 & 1367,960 & 20:34:41.80 & 60:08:54.9 & -9.44 & -0.31 & 0.32 & 0.63 & 9.9 \\
n6946-2210 & 1626,1003 & 20:34:35.34 & 60:09:03.4 & -10.09 & -0.86 & -0.04 & 0.11 & 24.7 \\
\hline
\end{tabular}
\end{table*}

\setcounter{table}{3}
\begin{table*}
\caption{(continued)}
\begin{tabular}{lrrrrrrrr} \\ \hline
  ID      &  $(x,y)$  & $\alpha$(2000.0) & $\delta$(2000.0) & $M_V$ & U-B & B-V  & V-I  & $R_e$ \\ \hline
n6946-4227 & 916,1625 & 20:34:53.25 & 60:10:58.1 & -8.99 & -0.37 & 0.10 & 0.32 & 11.0 \\
n6946-4238 & 124,1627 & 20:35:13.05 & 60:10:56.7 & -10.16 & -0.79 & 0.07 & 0.51 & 13.0 \\
n6946-4254 & 354,1630 & 20:35:07.30 & 60:10:57.8 & -9.82 & -0.78 & -0.16 & 0.29 & 10.4 \\
n6946-4309 & 199,1649 & 20:35:11.18 & 60:11:01.0 & -9.56 & -0.60 & -0.12 & 0.04 & 7.9 \\
n6946-4321 & 757,1652 & 20:34:57.23 & 60:11:02.8 & -9.18 & -0.08 & 0.27 & 0.41 & 5.3 \\
n6946-4338 & 102,1658 & 20:35:13.61 & 60:11:02.5 & -10.84 & -0.79 & -0.12 & 0.24 & 8.6 \\
\hline
\end{tabular}
\end{table*}
\end{document}